\documentclass[aps,prb,twocolumn,groupedaddress]{revtex4}
\usepackage{amssymb,amsmath,bm,dsfont}
\usepackage{epsfig,psfrag}
\usepackage{graphicx,subfigure}
\usepackage{wasysym}
\usepackage{color}
\usepackage[all]{xy}
\usepackage{hyphenat}
\usepackage{hyperref}		

\newcommand{\Hcal}{ \mathcal{H} }
\newcommand{\Ical}{ \mathcal{I} }
\newcommand{\tH}{ {\tilde{\mathcal{H}}} }
\newcommand{\tI}{ {\tilde{\mathcal{I}}} }
\newcommand{\kRev}{ \mathcal{P} }

\newcommand{\Path}{ \mathfrak{P} }
\newcommand{\ColTwo}[2]{ \begin{array}{ @{} c @{} } #1 \\ #2 \end{array} }

\newcommand{\SmallMxTwo}[2]{ \left[\! \begin{smallmatrix} #1 \\ #2 \end{smallmatrix} \!\right] }

\newcommand{\Tr}{ \operatorname{Tr} }

\newcommand{\coker}{ \operatorname{coker} }
\newcommand{\img}{ \operatorname{img} }

\newcommand{\phd}{ {\vphantom{\dag}} }		
\newcommand{\vk}{ {\mathbf{k}} }

\newcommand{\ZZ}{ {\mathbb{Z}} }

\begin{document}
\title{Quantized Response and Topology of Insulators with Inversion Symmetry}
\author{Ari M. Turner, Yi Zhang, Roger S.~K. Mong, Ashvin Vishwanath}
 \affiliation{Department of Physics,
University of California, Berkeley, California 94720, USA}
\date{\today}

\begin{abstract}
We study three dimensional insulators with inversion symmetry, in which other point group symmetries, such as time reversal, are generically absent.
Their band topology is found to be classified by the parities of occupied
states at time reversal invariant momenta (TRIM parities), and by three Chern
numbers. The TRIM parities of
any insulator must satisfy a constraint: 
their product must be $+1$. The TRIM parities also constrain the Chern numbers modulo two. When the Chern numbers vanish, 
a magneto-electric response parameterized by $\theta$ is defined and 
is quantized to $\theta=0,\,\pi$. 
Its value is entirely determined by the TRIM parities. These results may be useful
in the search for magnetic topological insulators with large $\theta$.
A classification of inversion
symmetric insulators is also given for general dimensions. 
 An alternate geometrical derivation of our results is obtained by 
using the entanglement spectrum of the ground state wave-function. 
\end{abstract}
\maketitle
\date{\today}

\section{Introduction}
Recently, the discovery of a class of insulator with nontrivial band
topology protected by time reversal symmetry\cite{topinsulators} has led to a range
of discoveries on the qualitative properties of solids. Soon after, it was
realized that this is part of a broader classification of gapped topological
phases with free fermions. If one includes particle hole symmetry, natural for
superconducting Hamiltonians, one ends up with ten symmetry classes\cite{ten,eleven} that
include both insulators and gapped superconductors. The symmetry transformations,
time reversal and particle hole symmetry, can be consistently generalized to
include disorder and inhomogeneities. Topological phases in these classes are
always characterized by protected surface states. However, one can also discuss
band topology in the presence of the point group symmetry of a crystal. That is,
 which classes of Hamiltonians can be smoothly connected to one another while
preserving the insulating gap and the symmetry? Although these distinctions require
considering a perfect crystal, and inevitably fade in the presence of
disorder, one often deals with crystals that are sufficiently
clean for such distinctions to be useful.
 
The general classification is a daunting task, given the vast number of symmetry groups of crystals. For example, there are 230 space
groups; if one adds transformation under time reversal symmetry to describe magnetic insulators, one has 1,651 different groups. 
We focus here on a simple case, 
inversion ($\mathbf{r}\rightarrow-\mathbf{r}$), a symmetry that is commonly realized 
in magnetic insulators. For example, all Bravais lattices are inversion 
symmetric. An additional virtue is that inversion symmetry
forces the magnetoelectric 
response to be isotropic and
quantized, allowing us to calculate it by qualitative methods.
Here, we will classify the topologically distinct inversion symmetric insulators, and determine their protected properties. (Phases of
antiferromagnets can also be classified using a similar approach\cite{Mong}.)

The classes of three dimensional inversion symmetric
insulators are found to be parameterized by
three Chern numbers and a set of
inversion parities. The Chern numbers, which are already present
in the absence of inversion symmetry, determine the  
quantized Hall response.
While this is an integer in two dimensions, it is given by a reciprocal lattice
vector $\tilde{\mathbf{G}}_{H}$ in a three dimensional crystal. 
Inversion parities are the only additional parameters that
appear from adding the extra symmetry. They
are defined at those momenta that are left invariant under
$\bm{\kappa} \rightarrow -\bm{\kappa}$, up to a reciprocal lattice vector.
This defines eight points in the 3D Brillouin Zone - the Time Reversal
Invariant Momenta (TRIM). States at these momenta can be classified by their
parities, eigenvalues under inversion, which are $\pm 1$. The set of inversion
parities of the filled levels at the TRIMs and $\tilde{\mathbf{G}}_{H}$ 
completely
specifies the topological class of the insulator.

Usually, distinct topological classes can be distinguished by their surface
state content. A different approach is to identify a quantized response, such
as Hall conductivity, that can distinguish topological classes. For inversion
symmetric insulators, in general, no protected surface states arise. This is
due to the fact that although inversion is present in the bulk, it is absent
at the surface of the system, which separates an inside from an outside.
However, quantized physical response functions exist, and are
protected by  inversion symmetry. These include the
magnetoelectric susceptibility
$\theta$, which is defined, for example, as the
polarization ($\bf P$) induced by a magnetic field ($\bf B$), in a parallel
direction: ${\bf P} = \theta \frac{e^2}{2\pi h} {\bf B}$. The
coefficient $\theta$ requires both bulk and surface to be insulating and is defined modulo $2\pi$.
Under time reversal,
$\theta \rightarrow -\theta$. Apart from the trivial solution $\theta=0$,
the ambiguity in the definition of $\theta$ allows also for $\theta=\pi$. The
latter was shown to occur in time reversal symmetric topological
insulators, when magnetic perturbations restricted to the surface open a
surface gap \cite{Shen,pi,pi2}. 

Under inversion symmetry, $\theta$ transforms in the same way as under
time reversal. Once again we expect $\theta=0,\,\pi$. The latter case
is particularly interesting since the magnetoelectric susceptibility
is large and no extra
effort will usually be required to gap out surface states, in contrast to
topological insulators. Typically, one does not expect surface modes in the
inversion symmetric case. The response $\theta$ is determined
just by the parities of the states at TRIMs, according to a formula
derived below (which is analogous
to formulae for the time reversal-symmetric cases 
\cite{TI,pi, symmetry-integral}).
Thus it will be possible to determine the
magnetoelectric susceptibility
without doing any band integrals. This should be useful in the search for
large $\theta$ insulators in magnetically ordered materials. Moreover, in
magnetic insulators a large spin orbit coupling is not a priori required to
obtain the phase with $\theta=\pi$, 
which opens up a wider selection of candidates.

A second set of results are in the form of constraints on the allowed values of
TRIM parities and Chern numbers $\tilde{\mathbf{G}}_H$. 
We first show that in any
insulator, a {\em parity constraint} is present. The product of the TRIM
parities of filled states is always $+1$. While this is a trivial condition
when time reversal is also present, and energy levels  come in Kramers pairs,
it is rather non-trivial for magnetic insulators. For example, the parity
assignment in Fig. \ref{fig:bingo}a 
rules out a band gap. A corollary of
 this result is that it is impossible to make a direct continuous transition
between a $\theta=0$ and $\theta=\pi$ insulator, if time reversal is broken.
One must either evolve through a non-insulating state, or a first order
transition. A continuous transition only exists if both time reversal and
inversion are present\cite{murakami}. Another constraint fixes
whether the three components of $\tilde{\mathrm{G}}_{H}$, 
are even or odd in relation to the TRIM parities. The intrinsic
polarization $\mathbf{P}$ of the insulator is also determined by the parities.

Finally we consider the entanglement.  The entanglement
spectrum of topological insulators contains
special modes\cite{lukasz,archiveAriFrankAV,otherentanglement}.
Section \ref{sec:entanglement} relates the number of gapless modes in
the entanglement spectrum
to the parity invariants. This result is used to rederive some of the electromagnetic properties in a simple fashion.

 The present article addresses
some questions left open in earlier work in which we participated.  Ref. 
\onlinecite{archiveAriFrankAV} discusses the entanglement spectrum of 
inversion symmetric insulators, without presenting the exact
relation to TRIM parities.
The proofs given here also complete the
reasoning presented in \cite{wanetal.} which studied the electronic
structure of a specific material using the
expressions for $\theta$ in terms of TRIM parities and the parity constraint.

In the next section, we give a brief survey of all of our results and a guide to the sections where their proofs can be found.

\section{Summary of Results}



The phases associated with a given symmetry can be
classified using some topology.
Each state can be classified by integer-valued parameters (or by integers
relative some modulus).
 These cannot change continuously,
so they determine phases.

Let us first define some conventions about the crystal lattice.  
We will for simplicity
assume that the lattice is cubic (although there is
no symmetry beyond inversion) and has
a lattice spacing equal to one unit.  All quantities will be written
with respect to a coordinate system $xyz$ that is aligned with the
axes of the crystal.
The results may
be translated
to systems with a \emph{general} Bravais lattice, by interpreting the
expressions in the right coordinate system.
Let $\mathbf{R}_i$
be the primitive vectors of the lattice and
let $\mathbf{g}^i$ be the reciprocal vectors, 
$\mathbf{g}^i\cdot\mathbf{R}_j=2\pi\delta^i_{j}$.  
If a vector is in real space, the coordinates $v^x,v^y,v^z$
refer to
$\mathbf{v}= v^x\mathbf{R}_1+v^y\mathbf{R}_2+v^z \mathbf{R}_3$.
Vectors $\mathbf{u}$ in reciprocal space should be expanded
in terms of reciprocal vectors,
$\mathbf{u}=\frac{1}{2\pi}(u_x\mathbf{g}^1+u_y\mathbf{g}^2+u_z\mathbf{g}^3)$. 
Electrical polarization is a real space vector; 
3D Hall conductivity and momentum
are in reciprocal space (and we also use upper and lower indices on
the coordinates
as a reminder
of what basis to use).

To classify bulk insulators, it is useful to
look at inversion parities, as in the study of spectra
of small molecules. The main difference is
that in solids, the occupied states can be labelled by momentum. Let these
states be given
by $\psi_{i\mathbf{k}}(\mathbf{r})=u_{i\mathbf{k}}(\mathbf{r})e^{i\mathbf{k}\cdot\mathbf{r}}$. The parity classification is only useful at
the ``TRIMs," the momenta
given by
\begin{equation}
\bm{\kappa}=\frac{n_1}{2}\mathbf{g}_1+\frac{n_2}{2}\mathbf{g}_2+
\frac{n_3}{2}\mathbf{g}_3
\end{equation}
where $n_1,n_2,n_3$ are integers.
Such a momentum maps to itself under inversion symmetry modulo the
reciprocal lattice, $-\bm{\kappa}\equiv\bm{\kappa}$.
Hence the wave functions at $\bm{\kappa}$ must be invariant,
and their parities can be defined:
\begin{equation}
\mathcal{I}\psi_{a\bm{\kappa}}(\mathbf{r})=\eta_{a}(\bm{\kappa})\psi_{a\bm{\kappa}}(\mathbf{r})
\end{equation}
Appendix \ref{app:q7} explains how to find these parities using a tight-binding
model.

We now introduce a key quantity $n_o(\bm{\kappa})$ at every TRIM $\bm{\kappa}$. This is defined as the \emph{number of states with odd parities} at that TRIM. 
Note, these cannot change without a phase transition 
(at least in a non-interacting system). Besides these 8 integers, the 
quantum Hall conductance gives three more invariant integers,
since it is quantized:
$\mathbf{G}_H=\frac{e^2}{2\pi h}\tilde{\mathbf{G}}_H$ where
$\frac{\tilde{\mathbf{G}}_H}{2\pi}$ has
integer components (according to the conventions defined above).

These $11$ integers, together with the total number of occupied bands $n$,
are the only parameters necessary to determine a phase--any two band structures with
the same integers can be tuned into one another without a phase
transition. This scheme is derived in Sec. \ref{sec:class}. Appendix \ref{app:americancheese} gives an alternative method that is easier to generalize.
These integers cannot be chosen independently of one another;
there are some relationships between their parities. 
These relationships can help
predict material properties, since the parities are easy to determine from the band structure.

\emph{Total Parity Constraint:}.
All the constraints
can be written in terms of the net parities:
\begin{equation}
\eta_{\bm{\kappa}}=(-1)^{n_o(\bm{\kappa})}=\prod_a \eta_a(\bm{\kappa}).
\label{eq:net}
\end{equation}
For any insulator, one can show:
\begin{equation}
\prod_{\bm{\kappa}}\eta_{\bm{\kappa}}=1.
\label{eq:constraint}
\end{equation}
That is, the total number of filled odd parity states must be even. This is shown in Sec. \ref{sec:constraint}.

\begin{figure}
\includegraphics[width=.45\textwidth]{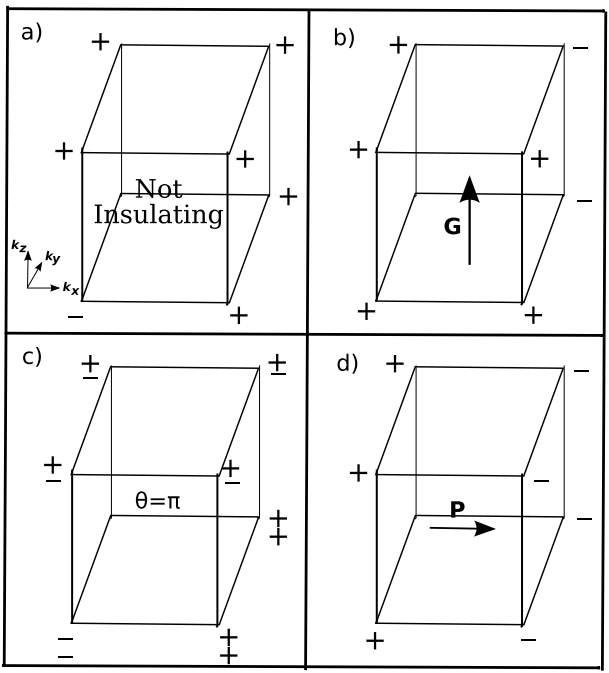}
\caption{\label{fig:bingo} Determining properties of systems using
parities.  The boxes represent an eighth of the Brillouin Zone; the TRIMs are at the corners. The signs represent the parities of the occupied states at the TRIMs. In (a) the parity constraint of even number of odd parity states is violated, hence it cannot be an insulator. In (b) the parities require a nonvanishing Hall conductance, with odd Chern number in the $k_x k_y$ planes. (c) Quantized magnetoelectric response $\theta=\pi$ determined from number of odd parity states being 2 (mod 4). (d) A parity configuration corresponding to a frozen polarization.}
\end{figure}

If a system has the parities in Fig. \ref{fig:bingo}a,
it must be metallic, because the parities do not satisfy
Eq. (\ref{eq:constraint}).  The gap must close at some momentum
$\mathbf{k}$ in the Brillouin zone. The simplest case
is if the system is a semimetal, with a single
pair of cone-points at the Fermi energy.

\emph{Quantum Hall Effect}.
Furthermore, the parities determine the quantum
Hall integers modulo 2.
The parity of $\frac{\tilde{\mathbf{G}}_{Hi}}{2\pi}$ is
constrained; e.g. the $z$-component satisfies
\begin{equation}
e^{\frac{i}{2}\tilde{G}_{Hz}}
=\prod_{\substack{\bm{\kappa};\\ \bm{\kappa}\cdot\mathbf{R}_z=0}} 
\eta_{\bm{\kappa}}
\label{eq:narrowhall}
\end{equation}
That is, whether the Hall conductivity along the $z$-direction is an
even or odd multiple of $2\pi$  can be determined by
multiplying the $\eta$'s around either of the squares parallel to the
$xy$-plane.
This result is derived in Sec. \ref{sec:polandqh}.

If a system has the parities shown in Fig. \ref{fig:bingo}b,
the Hall conductivity cannot vanish.  The component along the $z$-direction,
$G_{Hz}$, must be an odd multiple of $\frac{e^2}{hc}$ (per layer
of the crystal).

Eqs. (\ref{eq:constraint}),(\ref{eq:narrowhall}) are the
only relationships between the invariants--if the relationships
are satisfied, the invariants can be realized in principle.

\emph{Magnetoelectric Effect}.
The most interesting electromagnetic response that we discuss
is the magnetoelectric polarizability $\alpha^i_j$.
An applied magnetic field
induces a polarization, $P^i=\alpha^i_jB^j$. In
the absence of the quantum Hall effect,
$\alpha^i_j$ is well-defined. (Otherwise the polarization
can be neutralized by a flow of charge in the surface states associated
with the Hall effect.) 
The polarizability $\alpha^i_j$ is
odd under inversion:
Under inversion symmetry,
$\mathbf{P}$ changes sign, while $\mathbf{B}$ does not.

If the crystal is inversion symmetric, it seems that $\alpha$ must vanish.
However, $\alpha$ is ambiguous like the polarization.
An isotropic portion $(\frac{e^2}{h}\delta_{ij}\times\mathrm{integer})$ 
is indeterminate. (The magnetoelectric effect
affects only properties of the surface of a crystal.  An integer
magnetoelectric effect can be mimicked by a quantum Hall 
coating on the surface.)
Thus $\alpha^i_j$ can be isotropic,
$\frac{e^2}{2\pi h}\theta\delta^i_j$, if $\theta$ is a multiple
of $\pi$. 
To determine $\theta$, the $\eta_{\bm{\kappa}}$ parameters
are not sufficient, and we must return to the $n_o$'s:
\begin{equation}
\frac{\theta}{\pi}\equiv\frac{1}{2}\sum_{\bm{\kappa}} n_o(\bm{\kappa}) \pmod{2}.
\label{eq:magnetoelectric}
\end{equation}
According to Eq. (\ref{eq:constraint}), this is always an integer.
This expression is proved in Sec. \ref{sec:magnetoelectric}.

Fig. \ref{fig:bingo}c shows the parities of a model with 2 filled bands, which has a nontrivial $\theta$.  The total number of odd states at all the TRIMs is twice an odd number, hence $\theta$ is $\pi$. 
One interpretation
of $\theta=\pi$ is
that the insulator has a half-integer quantum-Hall effect on the
surface.  
Two filled bands are necessary for a nontrivial $\theta$.
The requirement $\mathbf{G}_H=0$ implies, for one filled band,
that the number of states is a
multiple of $4$. (See Eq. (\ref{eq:weaves} in the next section.)

\emph{Frozen Polarization}.
Finally, let us discuss what information is contained in the ``net parities"
$\eta_{\bm{\kappa}}$. Eq. (\ref{eq:narrowhall}) shows that
they determine the Chern numbers modulo 2. This accounts for
three of the eight parities.  Note that any pattern of parities
satisfying the constraint Eq. (\ref{eq:constraint})
can be factored into 7 basis patterns:
\begin{equation}
\eta(\bm{\kappa})=\pm (-1)^{\frac{1}{2\pi^3} (\tilde{G}_{Hx}\kappa_y\kappa_z+
\tilde{G}_{Hy}\kappa_x\kappa_z+\tilde{G}_{Hz}\kappa_x\kappa_y)}(-1)^{\frac{2}{\pi}\tilde{\mathbf{P}}_e\cdot\bm{\kappa}},
\label{eq:weaves}
\end{equation}
where the components of $\tilde{\mathbf{P}}_e$ 
are half-integers and the components
of $\tilde{\mathbf{G}}_H$ are integers times $2\pi$.  (The
factors of $\pi$ are added to make the exponents into integers.)
This equation agrees with Eq. (\ref{eq:narrowhall}), so
$\tilde{\mathbf{G}}_H$ is the Hall conductivity modulo 2.
When $\tilde{\mathbf{G}}_H=0$,$\eta_{\bm{\kappa}}$
varies as a plane wave on the vertices of the cube.
The components of $\tilde{\mathbf{P}}_e$ determine
the intrinsic polarization.

A crystal may have an intrinsic polarization 
when the Hall conductivity is zero\cite{restavanderbilt};
in this case, $\eta_{\bm{\kappa}}$
looks like a plane wave on the vertices of the cube
(see Fig. \ref{fig:bingo}d) and
$\tilde{\mathbf{P}}_e$ is the wave-number.
The spontaneous electrical polarization is similar to $\theta$; it 
is defined only\cite{restavanderbilt}
modulo a lattice vector times $e$. Inversion symmetry
constrains the components to be integers or half-integers times $e$.
Hence the polarization is determined by three bits 
and Eq. (\ref{eq:weaves}) gives all the information
about the polarization that can be obtained from bulk properties.
To get the polarization correct, one needs to include
the offset between the electrons 
and the compensating charges in the nuclei:
\begin{equation}
\mathbf{P}=e\tilde{\mathbf{P}}_e-\sum_i Z_i e\mathbf{r}_{Ni},
\label{eq:polarization}
\end{equation}
where $\mathbf{r}_{Ni}$
is the position vector of the $i^\mathrm{th}$ nucleus, with
charge $-Z_ie$. 
This result is derived in Sec. \ref{sec:polandqh}.

Consider the polarization of the crystal with the band structure illustrated
in Fig. \ref{fig:bingo}d; it is $\frac{e}{2}\mathbf{R}_1$
if the nuclei are all on the sites of the Bravais lattice.
This intrinsic dipole moment may correspond to actual ferroelectricity;
the crystal would have a surface charge and
a large electric field.  Alternatively,
the translational symmetry
of the surface may be spontaneously broken or the surface
may be metallic (see Ref. \onlinecite{v-andk-s} which
is summarized in appendix \ref{app:polarization}).

\emph{Parity constraints in general dimensions}.
The results in higher dimensions have a surprising
feature: as the number of dimensions increases
the sum of the $n_o$'s must be divisible by larger
and larger powers of $2$.

Specifically,
in $2s$-dimensions, the sum of the $n_o$'s is a multiple
of $2^{s-1}$. This multiple is
related to the $2s$-dimensional Chern number 
$\tilde{G}_{2s}$ (defined as a multiple of $2\pi$):
\begin{align}
	\frac{1}{2^{s-1}} \sum_{\textrm{TRIM }\bm\kappa} n_o(\bm\kappa)
		\equiv \frac{ \textrm{$s$\textsuperscript{th} Chern number} }{2\pi} \pmod{2} ;
\end{align}
the quantum Hall conductance is the 2-dimensional special case.

In $2s+1$-dimensions, the sum of the $n_o$'s is a multiple
of $2^s$ and is
related to the Chern-Simons integral
\begin{align}
	\frac{\theta_{2s+1}}{\pi} = \frac{1}{2^s} \sum_{\bm\kappa} n_o(\bm\kappa) \pmod{2} .
\end{align}
where the polarization and magnetoelectric effect are the one- and
three-dimensional versions.

Note that insulators with inversion symmetry are quite
different from ones:  There is
an insulator in $2s$ dimensions with 
a Chern number $\tilde{G}_{2s}$ equal to $1$ which has just $s$ filled bands\footnote{This follows from the relation between topological insulators
and the homotopy groups of Grassmann spaces $\mathcal{G}_{n,m}$;
the essential fact is that the homotopy group $\pi_{2s}(\mathcal{G}_{s,N-s})=\mathbb{Z}$ when $N$ is sufficiently large\cite{algebraictopology}.}
This insulator is not inversion symmetric, though.  The simplest
inversion symmetric insulator with the identical Chern number has a
minimum of $2^s$ bands, exponentially more bands than
are necesary without symmetry.

\emph{The Entanglement Spectrum}. An insulator with inversion
symmetry has a particle-hole symmetry $\mathcal{I}_e$ in
its entanglement spectrum $\epsilon_a(\mathbf{k})$ when it is cut
on a plane through a center of inversion. This makes it very easy to determine
qualitative properties of the Fermi arcs of the entanglement spectrum--it 
is possible to count (without topological arguments) the number
of zero modes in the entanglement spectrum
at the TRIMs $\bm{\kappa}_\perp$ along the surface. Let
$\Delta N_e(\bm{\kappa}_\perp)=\mathrm{tr}_{\epsilon=0}\mathcal{I}_e$;
that is $\Delta N_e(\bm{\kappa}_\perp)$ is the number of even modes minus
the number of odd modes with zero entanglement-energy at $\bm{\kappa}_\perp$. 

This can be expressed in terms of parities
of the bulk states through $n_o(\bm{\kappa})$. Define these parities
relative to an inversion center on the plane of the cut.
The quantity that
appears is $\Delta N(\bm{\kappa})=\mathrm{tr}_{E<0}\mathcal{I}$, or
$n-2n_o(\bm{\kappa})$:
\begin{equation}
\Delta N_e(\bm{\kappa}_\perp)=\frac{1}{2}(\Delta N(\bm{\kappa}_1)+\Delta N(\bm{\kappa}_2)),
\label{eq:firsthalf}
\end{equation}
where $\bm{\kappa}_1$ and $\bm{\kappa}_2$ are the two TRIMs that project
to $\bm{\kappa}_\perp$. In words: the difference
between the number of even and odd states on the entanglement ``Fermi
surface" at a TRIM is half the difference between the even
and odd states in the bulk, at the corresponding TRIMs.

%
%

\begin{figure}
\includegraphics[width=.45\textwidth]{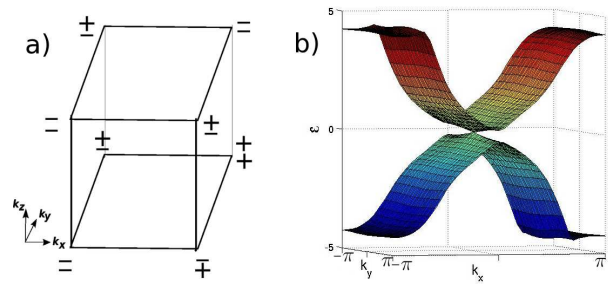}
\caption{\label{fig:clothchair}
Entanglement spectrum of a hopping Hamiltonian. a) The
parities at the TRIMs. b) The entanglement
modes on a cut parallel to the $xy$-plane.  Note that
there are two zero-modes at the TRIM $(0,0)$ and none at the other TRIMs,
as expected from the parities.}
\end{figure}

To illustrate an actual entanglement spectrum, we constructed
a Hamiltonian with a cubic unit cell whose inversion parities suggest
that $\theta=\pi$ and $\tilde{G}_{Hi}\equiv 0\ (\mathrm{mod\ 2})$.
The parities and the spectrum
are shown in Fig. \ref{fig:clothchair}. (The Hamiltonian is described
in Appendix \ref{app:q7}.)
The entanglement spectrum was calculated for a cut along the $xy$-plane.
As expected, there is a Dirac point at $(0,0)$.  In this case, there
are no chiral modes.

From the relation between the entanglement spectrum
and the parities, one can give alternative
derivations of some of the  results described above. This
formula also leads to a simple
determination of Fu and Kane's formula for the indices of
topological insulators.
These indices
describe the number (modulo 2) of physical surface states on a line
connecting two TRIMs, and these states are easy to count 
for the entanglement spectrum.
The entanglement spectrum can be continuously deformed into
the physical spectrum so it has the same index.

\section{Electromagnetic Evaluations and Equivalence of Insulators}

\subsection{Classifying Inversion Symmetric Insulators\label{sec:class}}
This section presents the classification of noninteracting
insulators with inversion symmetry. Hamiltonians
with inversion symmetry can be related to Hamiltonians
without any symmetry, which are already classified by Chern numbers.
The only new parameters occurring for inversion symmetry
come from the classification of "zero-dimensional insulators",
or finite molecules with inversion symmetry.

Consider the Hamiltonian $H(\mathbf{k})$ for the wavefunctions $\psi_{n\mathbf{k}}$.  This Hamiltonian
can be taken to be an $N\times N$ matrix by using
a tight-binding model with $N$ bands ($n$ of which are filled).
Since this Hamiltonian is inversion symmetric, 
\begin{equation}
I_0H(\mathbf{k})I_0=H(-\mathbf{k})
\label{eq:hi}
\end{equation}
where $I_0$ is the matrix describing how the orbitals within
the unit cell transform under inversion\footnote{There are
technical problems
when an orbital $|\alpha\rangle$ is at a half-lattice vector--choosing which copy
of it belongs to the unit cell breaks inversion symmetry.  The simplest
solution is to slice the orbital into two parts
that are images of each other under inversion, 
$|\alpha\rangle=\frac{1}{\sqrt{2}}(|\alpha_1\rangle\pm|\alpha_2\rangle)$.
The opposite combination of these orbitals is then assigned 
a very large energy so that it has no effect.}.

Assume that $H$ varies continuously without passing through a phase
transition.  This means that no states ever cross the
Fermi energy, $\mu=0$, say.
Let us determine when two Hamiltonians are in
the same phase--i.e., can be connected in this way.
We wish to find criteria on the
matrix fields $H(\mathbf{k})$ that can be used to determine which phase
they are in.

The only special points in the Brillouin zones are the
TRIMs, $\bm{\kappa}$, which are invariant under inversion symmetry.
Each of these points can be interpreted as
a zero-dimensional system, with a Hamiltonian $H(\bm{\kappa})$ 
that is invariant under $I_0$, since Eq. (\ref{eq:hi}) implies
$I_0H(\bm{\kappa})I_0=H(\bm{\kappa})$.
Let
$n_o(\bm{\kappa})$ be
the number of eigenvalues at negative energy which are odd under $I_0$.
The states at this TRIM can mix together, but even states can
mix only with even states and odd ones can mix only with odd ones, so the value
of $n_o(\bm{\kappa})$ cannot  change.

The Chern numbers of the Hamiltonian are topological winding
numbers which also turn out to describe the Hall conductivity\cite{TKNN,Avron}.
Because they are integers, they are also invariant.

We will now spend the rest of this section showing that
these give a complete classification of Hamiltonians with  inversion
symmetry.  That is, if $H(\mathbf{k})$ and $H'(\mathbf{k})$ 
are two Hamiltonians with the same number $n$ of occupied states,
and the same \emph{total} number of states $N$, and total number
of odd states $N_o$ (both filled and empty bands) and such that
\begin{align}\begin{split}
	&	n_o(\bm{\kappa})=n_o'(\bm{\kappa}) \,\, (\mathrm{for\ all\ TRIMs\ } \bm{\kappa} )		\\ 
	&	\tilde{\mathbf{G}}_H=\tilde{\mathbf{G}}_H',
	\label{eq:TheClassification}
\end{split}\end{align}
then (at least if $N-n,n\geq 2$) there is a family of 
Hamiltonians that connects $H(\mathbf{k})$ to $H'(\mathbf{k})$ 
without a phase transition and while satisfying
Eq. (\ref{eq:hi}).  (We may assume that $I_0$
remains constant since $N,N_o$ do not change.)

We do not usually consider the integers $N$ and $N_o$ to be important
invariants--their values can be changed by adding even or odd orbitals
with a very high energy.
In continuous
space, there are infinitely many available orbitals.

The assumption $N-n,n\geq 2$ is included because, when
there are too few bands,
there are some Hamiltonians
that cannot be deformed into one another
just because there are not enough
degrees of freedom.~\cite{joel}%
~\footnote{In a sequence such as $U(1), U(2)$, \textit{etc.}, the homotopy groups of the first few elements are irregular, but eventually stablizes farther in to the sequence.}
Our classification
theorem does not capture these distinctions, but the distinctions
are not related to any generic properties.
If one adds sufficiently many trivial occupied
and unoccupied bands to an insulator, any two
insulators with the same invariants can be deformed
into one another.

\textit{Proof for \eqref{eq:TheClassification}}:
Here we present the essential ideas in the classification of insulators.
App.~\ref{app:americancheese} gives a more systematic way of deriving the classification, including higher dimensions.

This result can be derived by relating a Hamiltonian
in $d$ dimensions to one in a smaller number of dimensions. 
Let us take $d$ to be arbitrary at first,
so that we can describe the general reasoning.
Let $\mathcal{H}_d$ 
be the space of general Hamiltonians in $d$ dimensions, while $\mathcal{I}_d$ 
is the subspace of Hamiltonians that also have inversion symmetry.
A generic Hamiltonian in $\mathcal{H}_d$ can be regarded as a closed
loop in $\mathcal{H}_{d-1}$:
For each value of $k_d$ (the $d^\mathrm{th}$ component of $\mathbf{k}$), 
consider the $d-1$-dimensional Hamiltonian
$H_{k_d}$ defined
by fixing one component of $\mathbf{k}$,
$H_{k_d}(k_1,k_2,\dots,k_{d-1}) \equiv H(k_1,k_2,\dots,k_d)$. This describes
a closed loop because the Brillouin zone is periodic.

\begin{figure}
\includegraphics[width=.45\textwidth]{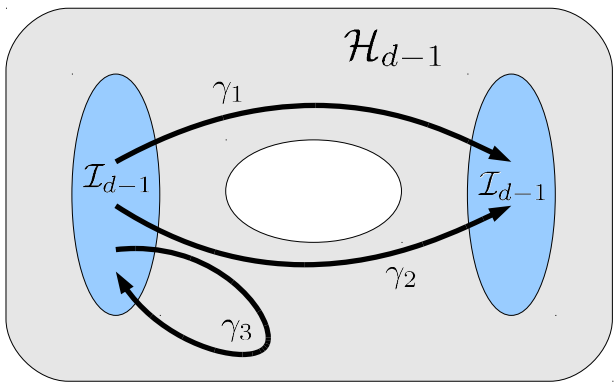}
\caption{\label{fig:manhole} Dimensional induction.
The grey region represents $\mathcal{H}_{d-1}$: each
point corresponds to a generic $d$-1-dimensional Hamiltonian.
The two ellipses on the side represent the components of
 $\mathcal{I}_{d-1}$, the
oHamiltonians with inversion symmetry.  Inversion-symmetric
$d$-dimensional Hamiltonians (three of which are shown)
are represented by arcs connecting
points in $\mathcal{I}_{d-1}$.
Two of these Hamiltonians are equivalent if the end-points
are in the same component of $\mathcal{I}_{d-1}$ and have the same
winding numbers around holes in the space (represented by the white ellipse).
For example, $\gamma_2$ and $\gamma_3$ are not equivalent
because their final end-points are in different components;
$\gamma_1$ and $\gamma_2$ are not equivalent because $\gamma_1\gamma_2^{-1}$
winds around the hole.
}
\end{figure}

A Hamiltonian in $\mathcal{I}_d$ is an arc in $\mathcal{H}_{d-1}$
with end-points in $\mathcal{I}_{d-1}$ (Fig. \ref{fig:manhole}) shows
$\mathcal{H}_d$ schematically, with
several possible arcs representing Hamiltonians. 
This arc is constructed by looking
at the cross-sections of $H(\mathbf{k})$ 
between $k_d=0$ and $k_d=\pi$.  The rest
of the Hamiltonian can be reconstructed using inversion symmetry.
The end-points have to be on $\mathcal{I}_{d-1}$ because the inversion
takes the $k_d=0,\pi$ cross-sections to themselves.

Thus, let us solve the following problem: consider
arcs $\gamma_1,\gamma_2$ in $\Hcal_{d-1}$ connecting two points in
the subspace $\Ical_{d-1}$.  What conditions 
ensure that it is possible to move
arc $\gamma_1$ to arc $\gamma_2$? This deformation is possible
if we can first slide the
end-points of $\gamma_1$ within $\mathcal{I}_{d-1}$ onto
the end-points of $\gamma_2$ 
and then smoothly deform 
the curves connecting them. Fig. \ref{fig:manhole} illustrates the problem.

We can thus classify d-dimensional Hamiltonians by solving two problems:
describing the different components
of $\mathcal{I}_{d-1}$, and classifying
the arcs connecting a pair
of points in $\mathcal{H}_{d-1}$ up to homotopy.

Let us now consider $d=3$.
The first step is analogous to the problem
we are trying to solve, just in one dimension less. 
(The components of $\mathcal{I}_{2}$ are just the different
classes of $2$-dimensional inversion-symmetric Hamiltonians.)
Let us suppose we know the solution to this problem,
so that the two arcs $\gamma_1,\gamma_2$
can be assumed to have the same end-points.

We now have to slide the interior of arc 1 onto arc 2.
Classifying arcs with fixed end-points in a given space
is closely related to classifying closed loops.
The \emph{loops} in $\mathcal{H}_2$ can be classified
by two winding numbers $\oint d\alpha(k_z)$ and $\oint d\beta(k_z)$,
where $\alpha$ and $\beta$ are angular variables around
holes in $\mathcal{H}_2$. This is a well-known
result in disguise.
Loops correspond to three-dimensional
Hamiltonians without any special symmetry, which
are classified (when there are sufficiently many bands) by
three Chern numbers.  The Chern number of the $k_z=0$
cross-section is a function of the points in $\mathcal{H}_2$,
not of the loops, and it is constant for the whole component we
are considering, so
$\tilde{G}_{Hz}$ does not matter. The remaining 
two Chern numbers are the winding numbers.

Now an arc connecting two points in $\mathcal{H}_2$ can wind around
a hole in the space any number of times just like a loop; it
just does not close on itself.  The topology
of loops is therefore determined by the value of $\int_0^{\pi} d\alpha(k_z)$
and $\int_0^{\pi} d\beta(k_z)$ as well.

The Chern numbers $\tilde{G}_{Hx}$ and $\tilde{G}_{Hy}$ of
the full Hamiltonian are given by 
$\oint_{-\pi}^\pi d\alpha(k_z),\oint_{-\pi}^\pi d\beta(k_z)$; 
hence the ``winding numbers" of the open arcs are half
as big as the Chern numbers, 
(by inversion symmetry).
So if the Chern numbers of the arcs 
are equal
the Hamiltonians are equivalent.

The only problem left is showing that the end-points can
be slid to one another under appropriate conditions.  
This is the same as classifying
inversion symmetric Hamiltonians in two-dimensions.
This problem may be reduced to one dimension.
The result is that, for
two Hamiltonians in $k_x-k_y$ space to be equivalent, the Chern 
numbers $\tilde{G}_{Hz}$
must be the same, and the one-dimensional boundary Hamiltonians
must be equivalent.

Now we must classify inversion-symmetric Hamiltonians in one dimension.
The problem reduces directly to zero
dimensions because all one dimensional Hamiltonians without symmetry
are equivalent.

Two zero-dimensional Hamiltonians (i.e., matrices!) are clearly
equivalent if the numbers of even and odd occupied states
are the same--just shift the energy eigenvalues so
that the two Hamiltonians match.  Hence the last 
condition is that the eight integers $n_o(\bm{\kappa})$
and the total number of occupied states $n$
must match.  (The original Hamiltonian
has bifurcated into eight zero-dimensional Hamiltonians
through the process of taking boundaries.)
The number of even and odd unoccupied states above the Fermi
energy must also be the same, but as mentioned above, there
are an infinite number of these in continuous space. 


Hence, three dimensional Hamiltonians are classified
by $\tilde{G}_{Hx},\tilde{G}_{Hy}$ and $\tilde{G}_{Hz}$
together with the parities at the TRIMs.

\subsection{Constraint on Parities\label{sec:constraint}}
Although the values of the 11 integers $n_o(\bm{\kappa})$ and $\tilde{\mathbf{G}}_{Hi}$
determine a phase, not all combinations of them are allowed.
First of all,
$\sum_{\bm{\kappa}} n_o(\bm{\kappa})$ must be even (Eq. (\ref{eq:constraint})).
There are two ways to see this.

For the first explanation, let us understand
a more general question:  consider a Hamiltonian (not
necessarily a gapped one) that is changing
as a function of time. 
We will try to understand what happens when the parities at
the TRIMs change.  The parities at a TRIM change only
when an even state at the TRIM below the Fermi energy
and an odd state above the Fermi energy (or vice versa) pass
through one another.
Appendix B shows that, each time
$n_o(\bm{\kappa})$ changes
by $1$ by means of such an interchange, 
a pair of Dirac points appears or disappears. Dirac points
are defined as points where the valence and conduction bands cross, with
a cone-shaped dispersion.  They are stable
unless they meet other Dirac points (with opposite ``chirality")
and annihilate.

Now start from a trivial Hamiltonian,
with all electrons glued to the Bravais
lattice; in this Hamiltonian, all the parities are even.
After
an odd number of changes of $n_o(\bm{\kappa})$, there are an odd
number of pairs of Dirac
points, so the crystal is not insulating.  If the system
is insulating $\sum_{\bm{\kappa}} n_o(\bm{\kappa})$ must
be even.

 The stability of the Dirac points
is explained in part by the basic
result on degeneracies of eigenvalues, rather than by symmetry:
in order to tune a Hamiltonian to a point where there is a degeneracy,
three parameters are sufficient. Since $H(\mathbf{k})$ 
is a function of three momenta, these may be tuned to a point where there
is a degeneracy, provided $H$ is close enough to having
a degeneracy in the first place.

\begin{figure*}
\includegraphics[width=\textwidth]{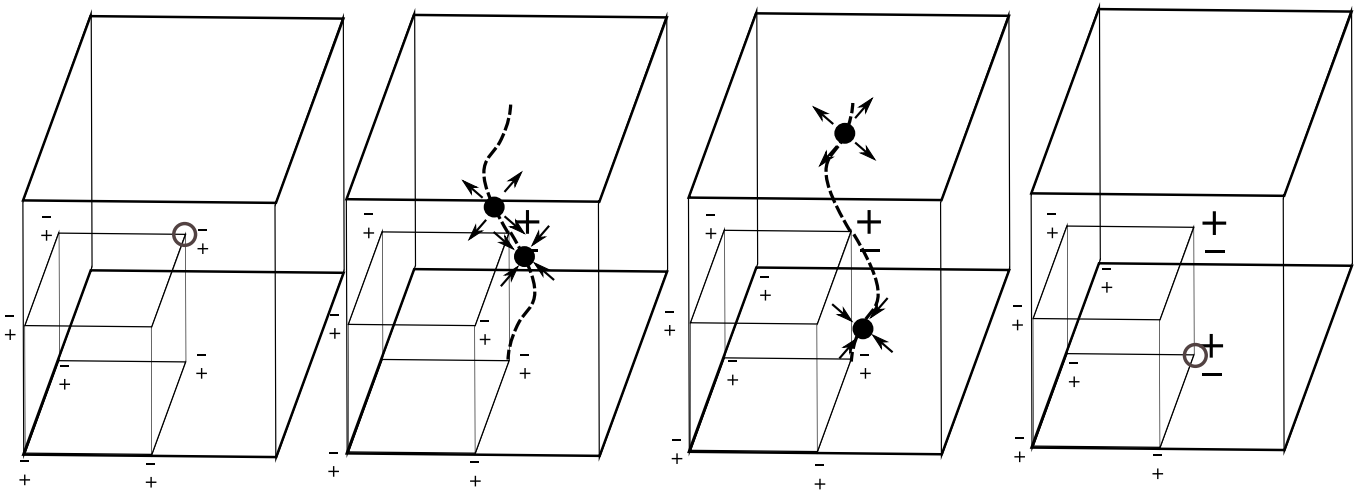}
\caption{\label{fig:lifecycle}Changing the parities of bands at two TRIMs.
The figures represent the Brillouin zone of a system with two bands,
one of which is filled.  Initially, all the filled bands
have parity $+1$, but the parities at two points are changed
with the assistance Dirac points which
also act as monopoles in the Berry flux (see appendix \ref{app:monopoles}). 
A pair of monopoles forms at one TRIM and  they
move to another TRIM where they annihilate.  In the process, the parities
of the states at both TRIMs are reversed. The open circles indicate
where the monopoles start out and disappear.
}
\end{figure*}

Any set of parities $n_o(\bm{\kappa})$ satisfying 
$\prod_{\bm{\kappa}}\eta_{\bm{\kappa}}=1$ can
be realized in an insulator. There is never a direct
phase transition (even with fine tuning) between two such phases when
two or more $\eta_{\bm{\kappa}}$'s flip sign.
When two modes
cross at one TRIM in order to change
the value of $n_o$ there, Dirac points will form, and the
system will be a semimetal.
The Dirac points must then move
to the second TRIM and
reannihilate, so that the system becomes an insulator again, as illustrated
in Fig. \ref{fig:lifecycle} 

The alternative derivation of the constraint studies the Bloch
eigenfunctions for a fixed Hamiltonian.
  Let us first suppose there
is a single occupied band $|\psi_{1\mathbf{k}}\rangle$.
To determine whether the wave function is even or odd, let us take
its overlap with an even orbital $|s\rangle$
centered on the origin. 
Define $s_1(\mathbf{k})=\langle s|\psi_{1\mathbf{k}}\rangle$.
Plot the solutions in the Brillouin zone to
\begin{equation}
s_1(\mathbf{k})=0
\end{equation}
This equation
is a complex equation, amounting to two equations
in three variables, so its solutions are curves.
At a TRIM, $\psi_{1\bm{\kappa}}(\mathbf{r})$ is 
either even or odd.  If it is odd, its overlap
with $|s\rangle$ vanishes.  Generically, the
converse is also true.
Hence,
there is one curve through each TRIM 
where $|\psi_{1\bm{\kappa}}\rangle$ is odd.

\begin{figure}
\includegraphics[width=.45\textwidth]{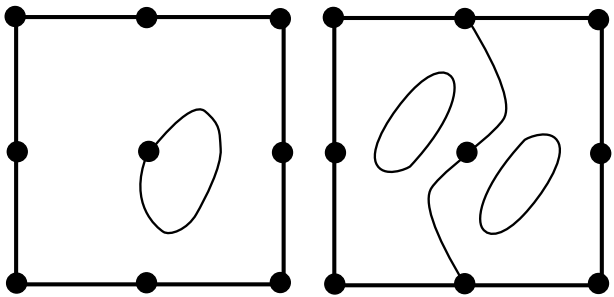}
\caption{\label{fig:syrup} Curves which are inversion symmetric
must pass through an even number of TRIMs. Left, an attempt
at drawing a curve that passes through one TRIM fails to
be inversion symmetric. Right, an inversion symmetric figure; if there
is just one curve passing through one of the TRIMs, it must go all the
way around the Brillouin zone and pass through another TRIM on the way.}
\end{figure}

But since the curves are inversion symmetric, they
must pass through an even number of TRIMs (see Fig. \ref{fig:syrup}). Hence, 
the total number of TRIMs where $\mathcal{I}\psi_{1\bm{\kappa}}=-\psi_{1\bm{\kappa}}$ 
is even.

When there are several filled bands which do
not touch each other, $\prod_{\bm{\kappa}} \eta_{a}(\bm{\kappa})=1$ for
each band separately, and $\prod_{\bm{\kappa}}\eta_{\bm{\kappa}}=1$
follows.
If bands do touch, another step is required (the product for
a single band may be $-1$, but the total will still
be $+1$). Consider the
curves determined by $s_a(\mathbf{k})=\langle s|\psi_{a\mathbf{k}}\rangle=0$ 
for all the occupied
bands $1\leq a\leq n$.  Some of these curves may be open arcs
because $s_a(\mathbf{k})$ becomes discontinuous
at Dirac points. Such arcs 
can only end at a Dirac point where the band touches
another band.  In
this case, there is also an arc leaving the Dirac point in the other band
(see App. \ref{app:monopoles}).
Putting all the arcs from the occupied states
together therefore produces a set of closed curves; these curves will
be variegated if the arcs in each band are imagined
to have different colors, but they are still closed. 
Hence we can still deduce that $\sum_{\bm{\kappa}} n_o(\bm{\kappa})$ is even.



\subsection{Polarization and Hall Conductivity\label{sec:polandqh}}
The constraint just derived is not the only constraint on the 11 invariants.
The parities of the $\tilde{\mathbf{G}}_H$ are determined by the parities
of the $n_o$'s (see Eq. (\ref{eq:narrowhall})).  
We will prove this momentarily, but
it is logically necessary to derive the expression for
the polarization first.
Both these quantities are related to the Berry connection,
a vector function in momentum space.
For a single band, the Berry connection is defined by 
\begin{equation}
\mathbf{A}_a(\mathbf{k})=i\langle u_{a\mathbf{k}}|\nabla_\mathbf{k}
|u_{a\mathbf{k}}\rangle
\label{eq:strawberry}
\end{equation}
and the total Berry connection $\mathbf{A}(\mathbf{k})$ is the sum
of the Berry connections of the occupied bands.

We will prove that the intrinsic polarization is given (modulo  $e$ times
a lattice vector) by Eqs. (\ref{eq:weaves}),(\ref{eq:polarization}).  
In general, the polarization per unit cell
is given by
\begin{equation}
\mathbf{P}=\mathbf{P}_e-\sum_i Z_i e(x_i-x_0).
\end{equation}
The second term is the polarization of the nuclei in the unit cell
relative
to an origin, $x_0$, and the first term is the polarization
of the electrons relative to $x_0$.  Since the electrons
are delocalized, calculating this contribution is
subtle.  It is given, according to Ref. \onlinecite{restavanderbilt}, by
\begin{equation}
P_e=\frac{e}{2\pi}\sum_n\int dk A^x(k).
\label{eq:resta}
\end{equation}

This expression for the
polarization is ambiguous up to multiples of $e$, as it should be,
on account of surface charge.  
If the unit cell is redefined, some nuclei locations are shifted
by one unit. Likewise if the Bloch wave functions are redefined
by $u_{nk}\rightarrow e^{i\theta(k)} u_{nk}$,
then the polarization shifts by $\frac{e}{2\pi}(\theta(2\pi)-\theta(0))$,
an integer multiple of $e$ if $e^{i\theta}$ winds around the
unit circle.

To evaluate the polarization of an insulator with inversion symmetry,
set $x_0=0$, the inversion center.
First, consider a single band.
The wavefunctions at $k$ and $-k$ must be the same
up to a phase, so $\psi_{-k}=e^{i\theta(k)}\psi_{k}$ for some
phase $\theta(k)$.
Therefore, $A_{k}+A_{-k}=\theta'(-k)$.  Combining $k$
and $-k$ together in Eq. (\ref{eq:resta}) leads to 
$\tilde{P}_e=\frac{1}{2\pi}\int_0^\pi \theta'(-k)dk=\frac{1}{2\pi}
(\theta(-\pi)-\theta(0))$. Now
$e^{i\theta(k)}$ is the parity of the wave function $\pm 1$
at TRIMs $k=0,\pi$. Hence if the parities
at the TRIMs are different then
the polarization is a half integer.
If there are many bands, we may sum the polarization
over all of them and we find in general that
\begin{equation}
(-1)^{2\tilde{P}_e}=\eta_0\eta_\pi.
\label{eq:seesaw}
\end{equation}

The result for the polarization in three dimensions
is derived from this below, but Fig. \ref{fig:bingo}d suggests  
the reason.
Each of the four vertical
lines through TRIMs looks like a one-dimensional insulator
with half-integer polarization, so the net polarization per unit
cell of the three-dimensional crystal is also
$\frac{e}{2}\mathbf{R}_3$.
Note that half of any Bravais lattice vector is an inversion center,
so there are eight inequivalent inversion centers.  The parities
depend on which center they are measured relative to, so $\mathbf{P}_e$
is origin-dependent.  But $\mathbf{P}$ is not because the contribution
from the nuclei depends on the origin in the same way.

Now let us consider the
Chern number for a \emph{two} dimensional system, $H(k_x,k_y)$,
and show that $(-1)^{\tilde{G}_{Hz}}=\prod_{\bm{\kappa}}\eta_{\bm{\kappa}}$. This
is the two-dimensional version of Eq. (\ref{eq:narrowhall}).
The Hamiltonian leads to a 1-D Hamiltonian $H_{k_y}$ when
$k_y$ is fixed.
As $k_y$ changes, the polarization $P(k_y)$ of the one dimensional
system changes.  This means current must flow from one end to the other.
According to Thouless's pumping argument,
the Hall conductivity $\tilde{G}_{Hz}$ is
equal to the total charge (divided by $e$) that flows
when $k_y$ changes by $2\pi$:
  (In real space, the one-dimensional system
is just the two-dimensional system rolled into a tube along the
$y$ direction, with period equal to a single cell.
Changing $k_y$ corresponds to applying an EMF around the $y$-direction.)

Thus $\tilde{G}_{Hz}=-\frac{1}{e}\int_{-\pi}^{\pi} dP(k_y)$.
(The polarization is not single-valued if $\tilde{G}_z\neq 0$.)
Now if
 $\prod_{\bm{\kappa}} \eta(\bm{\kappa})=-1$ (as in either of the $2D$
layers in Fig. \ref{fig:bingo} b) then the polarizations
at $k_y=0$ and $k_y=\pi$ differ by a half integer.  Thus,
$\int_0^\pi dP=(k+\frac{1}{2})e$.
By inversion symmetry,  $dP(k_y)=dP(-k_y)$ (since
$P(k_y+dk_y)-P(k_y)=-P(-k_y-dk_y)+P(-k_y)$).
Hence between $-\pi$ and $0$ the polarization changes by
the same half-integer, and the full change
in the polarization is odd, 
$\tilde{G}_{Hz}=-(2k+1)2\pi$.

This section concludes with the generalizations of these results to three dimensions.
The expression for the Hall coefficient
in three dimensions Eq. (\ref{eq:narrowhall}) is basically a restatement of the
two-dimensional result. Each of the three components
of $\mathbf{G}_H$ is equal to the two-dimensional
Hall coefficient
for any cross-section of the Brillouin zone:
\begin{eqnarray}
G_{Hz}&=&\int \frac{dk_z}{(2\pi)} G_{Hz}^{2d}(k_z)\nonumber\\
&=& G_{Hz}^{2d}(k_{z0})
\end{eqnarray}
since the Chern number for any cross-section $k_z=k_{z0}$
is the same since it is a topological invariant.

The Hall coefficient can be obtained
modulo 2 by looking at an inversion symmetric
plane, either $k_z=0$ or $k_z=\pi$, giving Eq. (\ref{eq:narrowhall}).
Note that this gives another reason for the constraint 
$\prod_{\bm{\kappa}} \eta_{\bm{\kappa}}=1$: the two planes have
to agree about $\tilde{G}_{Hz}$'s parity. 

The expression for
the Hall coefficient can be guessed by looking
at Fig. \ref{fig:lifecycle}. The initial phase is trivial.
When the monopoles
move through the Brillouin zone and annihilate, they leave behind a magnetic
flux, so the Chern number is $2\pi$. 
The flux's direction is parallel to the
edge where the parities of the occupied states have been flipped.

The polarization in three
dimensions is well-defined if $\mathbf{G}_H=0$; then
there are no surface modes.
According to Eq. \ref{eq:weaves},
the pattern of signs is then a plane wave, so all the $1d$
polarizations at TRIMs,
$P_{1d}^x(\kappa_y,\kappa_z)$, are the same. 
This value in fact 
coincides with the three dimensional
polarization. 

The three-dimensional polarization is the integral
over one-dimensional polarizations, $P^x=\iint\frac{dk_y dk_z}{(2\pi)^2} 
P_{1d}^x(k_y,k_z)$, if $G_{Hy}=G_{Hz}=0$.  (This condition ensures
that $P^x$ is single-valued.)
The integrand is not a constant, but by inversion
symmetry $P_{1d}^x(k_y,k_z)-P_{1d}^x(0,0)=-(P_{1d}^x(-k_y,-k_z)-
P_{1d}^x(0,0))$. (Intuitively, one expects $P_{1d}^x(k_y,k_z)$
to be an odd function of the wave number, but
only differences in polarization are inversion
symmetric because of the ambiguity in the polarization.)
Hence
$P^x=P_{1d}^x(0,0)$, which is given by Eq. (\ref{eq:weaves})
and Eq. (\ref{eq:polarization}).

Note that if $\mathbf{G}_H$ is nonzero, then part of the polarization
is still well-defined.  If $\mathbf{G}_H$ is parallel to $\mathbf{g}_1$ 
then \emph{this}
component of the
polarization cannot leak through the surface modes.
The chiral modes circle
around the $x$-axis, so they do not provide a short circuit
between the right and left faces of the crystal.
In fact, the same reasoning shows that $P_x$
is still given by Eqs. (\ref{eq:weaves}),(\ref{eq:polarization}).

\subsection{Magnetoelectric Response\label{sec:magnetoelectric}}

Now we justify the relation $\theta=\pi\frac{1}{2}\sum_{\bm{\kappa}} n_o(\bm{\kappa})$
(Eq. (\ref{eq:magnetoelectric}))
by using the classification of topological insulators to deform
an arbitrary inversion-symmetric Hamiltonian $H$ to one that also has
time-reversal symmetry.  Eq. (\ref{eq:magnetoelectric})
is known for Hamiltonians with time-reversal in addition
to inversion, either by directly evaluating
the expression for $\theta$ in terms of the
Berry connection \cite{symmetry-integral} or by
combining the results of \onlinecite{pi,pi2} and \onlinecite{TI}.

Hamiltonians with time-reversal symmetry and inversion symmetry
have extra constraints on the $\tilde{\mathbf{G}}_H$
and $n_o$'s.  The states at the TRIMs
come in Kramers doublets, so $n_o(\bm{\kappa})$ and $n$ are even, 
and the Hall conductivity must be $0$ by symmetry.
By the classification theory,
an inversion symmetric insulator can be deformed into one that
is also time-reversal symmetric if it has these properties, too, 
and Eq. (\ref{eq:magnetoelectric}) follows.

Now we are already assuming that the Hall conductivity is zero,
since the gapless surface modes can interfere with the 
magnetoelectric response. 
When some of the $n_o$'s are odd there is an additional
step before it is possible to deform $H$ 
into a time-reversal symmetric Hamiltonian.
We superimpose 
an independent inert system onto $H$ and
then deform the combined system.  
Consider an ``ionic" or ``frozen" insulator with protons
on the Bravais lattice and electrons fixed on certain sites, and with
all hopping amplitudes equal to 
zero.  This insulator does not contribute to $\theta$.  If the electrons
are at half of a Bravais
lattice vector $\mathbf{p}=\frac{\mathbf{R}}{2}$ and its translates, then
it is inversion symmetric.

Though this insulator does not contribute to $\theta$,
its parities depend on $\bm{\kappa}$.  This is enforced
by Eq. (\ref{eq:weaves}) since the insulator has a polarization.
Calculating these parities directly sheds light on Eq. (\ref{eq:weaves}).
The parities are obtained by
transforming to a basis of plane wave states,
\begin{equation}
|\mathbf{p}\rangle_\mathbf{k}=
\sum_{\mathbf{R}}e^{i\mathbf{k}\cdot(\mathbf{R}+\mathbf{p})}|\mathbf{R}+\mathbf{p}\rangle
\end{equation}
where $|\mathbf{r}\rangle$ represents an electron at an orbital
centered around $\mathbf{r}$.
When $\mathbf{k}$ is a TRIM,
$\mathcal{I}|\bm{\kappa}\rangle_\mathbf{p}=e^{2i\mathbf{p}\cdot
\bm{\kappa}}|\bm{\kappa}\rangle_\mathbf{p}$, assuming the orbitals
are even under inversion.

Now consider an arbitrary Hamiltonian $\mathcal{H}$ without
Hall conductivity.
Eq. (\ref{eq:weaves})
implies that the net parities $\eta(\bm{\kappa})$
vary just as they do in the ionic crystal (except possibly with
an overall sign).  Therefore, we can add a copy of an ionic
crystal with a polarization that matches the intrinsic
polarization of $\mathcal{H}$, leading
to an extended Hamiltonian $\mathcal{H}_{\mathrm{ext}}$. 
All the $\eta_{\bm{\kappa}}$'s
will become $\pm 1$.  This sign can be changed
to $+1$--redefine the orbitals of the frozen
crystal to be odd under inversion. (If \emph{all} the signs were \emph{initially}
$-1$ add a frozen crystal with $\mathbf{p}=0$ and odd parity orbitals.)

Now the $n_o(\bm{\kappa})$'s have become even.  If $n$
is odd then add an unpolarized crystal,
with a single electron in an even orbital on the origin and
its translates. Now both $n$ and $n_o(\bm{\kappa})$ are both
even.

This enlarged crystal's $\theta$ may be determined by
deforming it to a time-reversal symmetric insulator, so
$\frac{\theta_{\mathrm{ext}}}{\pi}\equiv 
\frac{1}{2}\sum_{\bm{\kappa}} n_{o,\mathrm{ext}}(\bm{\kappa})$.
But it has the same magnetoelectric
response $\theta_{\mathrm{ext}}$ as the original crystal, 
and the frozen crystals changed the number of odd states
at the TRIMs by a multiple of $4$; hence this formula
applies to the original crystal also.

\subsection{Other Quantized Responses?}
We have found that there are eight integers $n_o(\bm{\kappa})$
that are constant within a given phase. However, the calculations
so far have only found that a few mod. 2 combinations of them that have
interpretations as response functions. We think 
that there are no quantized responses besides $\theta$ and $\mathbf{G}_H$,
for any applied field or geometry.  For any
set of $n_o(\bm{\kappa})$'s, such that $\mathbf{G}_H$
and $\theta$ are zero,  it is possible
to find a Hamiltonian where the particles have zero hopping.
  
Let us start from a general vector of values $n_o(\bm{\kappa})$ and
try to build it out of the $n_o$ vectors for the
frozen (or ionic) insulators described in the previous section.
Frozen insulators do not have any response.
Regard
$n_o(\bm{\kappa})$ as an eight-dimensional vector of integers $\mathbf{n_o}$.
The vectors in this space can be pictured
by labelling the vertices of a cube with eight numbers.

The frozen insulators
described in the previous section give
eight vectors $\mathbf{f}_\mathbf{p}$
in this space, as $\mathbf{p}$ varies over all possible
polarizations.  ($\mathbf{f}_0$ represents the insulator where
each $n_o=1$.)
  Since $\mathbf{n_o}$ is eight-dimensional and
there are eight $\mathbf{f}_\mathbf{p}$'s,
every insulator's vector can be decomposed
in terms of these frozen vectors.  This suggests that every
insulator can be built out of frozen insulators.  This
is not really true though, because  the superposition may involve
non-integer coefficients. There is no way to
take a fractional multiple of a system. 

For each vector that \emph{can} be expanded as
$\mathbf{n_o}=\sum_{\mathbf{p}} r_{\mathbf{p}} \mathbf{f}_{\mathbf{p}}$ where
the $r_{\mathbf{p}}$'s are integers, there is a frozen
insulator with the same $\mathbf{n_o}$ up to a constant.
Hence these $\mathbf{n_o}$'s do not require nontrivial quantized
responses. 
For each $\mathbf{p}\neq 0$, combine together  
$|r_\mathbf{p}|$ copies of the insulator with the polarization 
$\tilde{\mathbf{P}}_e=\mathbf{p}$. (The parities of
the inert orbitals should be chosen based on the sign of $r_{\mathbf{p}}$.)

How close can one get to a get $\mathbf{n_o}$ can one get
with such a sum of frozen vectors?
Appendix \ref{app:inertia} shows that for each eight-component 
integer vector $\mathbf{n_o}$ one can subtract
a frozen vector
$\mathbf{f}$
so that
\begin{equation}
\mathbf{n_o}-\mathbf{f}=u_{xy}\mathbf{v}_{xy}+u_{xz}\mathbf{v}_{xz}+u_{yz}\mathbf{v}_{yz}+w_{xyz}\mathbf{v}_{xyz}\label{eq:uvwxyz}
\end{equation}
where the $u$'s are each $0$ or $1$ and $\mathbf{w}_{xyz}$ is $0,1,2$ or $3$.
  The $\mathbf{v}$'s are the following vectors. 
The first, $\mathbf{v}_{xy}$ corresponds
to the insulator illustrated in Fig. \ref{fig:bingo}b, with
the quantum Hall response. It
has ones on the \emph{edge} of the cube described
by $\kappa_x=\kappa_y=\pi$
and zeros everywhere else.  The other $v$'s with two labels
are defined symmetrically.
The vector $\mathbf{v}_{xyz}$
has a one at one \emph{corner} of the cube ($\kappa_x=\kappa_y=\kappa_z=\pi$).
If the vector $\mathbf{n_o}$ of an insulator is decomposed in this way, 
the constraint $\sum_{\bm{\kappa}} n_o({\bm{\kappa}})\equiv 0\ (\mathrm{mod\ }2)$ 
implies that $w_{xyz}$ is even (either $0$ or $2$ modulo $4$). 
Hence the difference betweena generic insulator and a frozen
one is characterized by values of four numbers--each $u$ can
be $0$ or $1$ and $w_{xyz}$ can be $0$ or $2$.
Thus any property of the responses that is contained in the
$\mathbf{n_o}$'s is determined by these four parameters. 

Eqs. \ref{eq:uvwxyz} and \ref{eq:narrowhall} show
that the $u$'s describe the quantum Hall conductivity
modulo 2, while $\theta=\pi$ if $\mathbf{G}_H=0$ and
$w_{xyz} \equiv 2 (\mathrm{\ mod\ }4)$.
There seem to be no other responses.

In spite of this, it is possible to measure some information about $\mathbf{f}$
by considering the polarization.  (The argument does not rule this out
because polarization is static.)

\section{Parities and the Entanglement Spectrum\label{sec:entanglement}}

The relations between the properties
of the insulator and the parities
can be derived geometrically using the entanglement
spectrum of the insulator; there is a rule for counting
the number of states in this spectrum based on the parities.

The entanglement spectrum is defined using the Schmidt decomposition.
The insulator is cut by an imaginary plane passing through a center of
inversion symmetry. The many-body ground state wave function
then decomposes,
\begin{equation}
|\Psi\rangle=\frac{1}{\sqrt{Z}}\sum_\alpha e^{-\frac{E^e_{\alpha}}{2}}|\alpha\rangle_L|\alpha\rangle_R\label{eq:schmidt}.
\end{equation}
where $Z$ is a normalization constant and $E^e_\alpha$ controls
the weight of a given term. (It is called the entanglement ``energy"
because this formula looks similar to the Boltzmann distribution. The states
can be called the entanglement states.)

When the wave function of the entire system is a Slater determinant,
the entanglement
states $|\alpha\rangle_L$ are Slater determinants, too\cite{Peschel}.  They
are formed by selecting wave functions from a special family of
single-particle wave functions
$f^L_{i\mathbf{k}_\perp}(\mathbf{r})$.  (These
states may be labelled by the momentum along the surface, $\mathbf{k}_\perp$,
by translational symmetry.) 
Each of these
wave-functions has an associated ``energy" $\epsilon_{Li}(\mathbf{k}_\perp)$
(as if they were eigenfunctions of a single-particle Hamiltonian).
The entanglement ``energy" $E^e_\alpha$ is the sum of all the ``energies"
of the occupied states.
The entanglement spectra, $\epsilon_{Li}(\mathbf{k}_\perp)$, can
be used to determine ``topological" properties of
the system, since these spectra may be continuously
deformed into the physical surface spectrum of the 
system\cite{lukasz,archiveAriFrankAV}.

Determining basic properties of the entanglement spectrum
is simple in the presence of inversion symmetry.
The entanglement spectrum has a particle-hole symmetry $\mathcal{I}_e$
that implies a rule
for finding the number of entanglement states at each surface
TRIM.
The $\mathcal{I}_e$ symmetry takes each mode
to another mode whose
momentum $\mathbf{k}_\perp$ and ``energy" $\epsilon_i(\mathbf{k}_\perp)$ 
have the opposite sign. Let us regard $0$ as the Fermi ``energy";
the state in the Schmidt decomposition with the smallest ``energy"
(i.e., the largest weight) is obtained by filling up all states
with $\epsilon<0$.

At surface TRIMs
$\mathcal{I}_e$ ensures that
states appear in pairs with
energies $\pm \epsilon$ when $\epsilon\neq 0$.  There can be
a single mode at zero. This mode will stay
exactly at zero no matter how
the system is changed, because moving away would break the symmetry.
More can be said about the zero-energy states:
the ``index" at each TRIM can be determined; this is
the difference $\Delta N_e(\bm{\kappa}_\perp)$
between the number of modes of even and odd parity.
This quantity is significant because even and odd zero-energy
states can ``cancel" one another and move to nonzero energies
$\pm\epsilon$, while two states of the same parity cannot.
If two states
evolve into the eigenstates $f_1$ and
$f_2$ with energies of opposite sign,
then $f_1$ and $f_2$ are orthogonal states exchanged by $\mathcal{I}_e$.
The corresponding parity eigenstates, 
$\frac{1}{\sqrt{2}}(f_1\pm f_2)$, have
\emph{opposite}
parities. 
 
The imbalance number can be found directly from the bulk
band structure, 
\begin{equation}
\Delta N_e(\bm{\kappa}_\perp)=1/2(\Delta N(\bm{\kappa}_1)+\Delta N(\bm{\kappa}_2))
\label{eq:half}
\end{equation}
where $\bm{\kappa}_1$ and $\bm{\kappa}_2$ are the two bulk inversion-symmetric
momenta that project to $\bm{\kappa}_\perp$ and
$\Delta N(\bm{\kappa}_1)$ (e.g.) is related to $n_o(\bm{\kappa}_1)$;
it is the difference between
the number of even and odd occupied states at $\bm{\kappa}_1$,
that is $n-2n_o(\bm{\kappa}_1)$. In one dimension,
for example, there is only one $\Delta N_e$ (since there is
only one surface momentum) to determine, and $\Delta N_e$ \emph{is
equal to the number of even states at $\kappa=\pi$ minus the
number of odd states at $\kappa=0$}. 
The parities
of the bulk states are to be calculated relative to
an inversion center that is on the cutting plane. (Parities 
depend on the inversion center, as does the entanglement spectrum.  The
physical responses described above do not\footnote{Suppose in particular
that we are working with a hopping model consisting
of a chain of identical sites each with multiple orbitals.
We have to make a cut on a \emph{bond}
midway between two sites, while defining
inversion parity relative to \emph{sites} is the most natural choice.
Defining the parities this way, $\Delta N_e=n_o(\pi)-n_o(0)$
since the parities of states at $\kappa=\pi$
are switched by the change in inversion center from a bond
to a site. This is equivalent to the
result obtained in Sec. IIC of Ref. \cite{Hughesentanglement}.
The factor of 2
in that article is present because the authors
consider periodic boundary conditions,
and cutting the chain creates two ends.}

The derivation of this requires some results of Refs.
\onlinecite{Peschel,boteroklich}
whose derivations are summarized in appendix \ref{app:entanglement}. 
As the structure of the entanglement spectrum suggests,
the entanglement modes are actually eigenfunctions of a Hamiltonian
$H_L$ defined on the part of space to the left of the cutting plane.
The eigenvalues of $H_L$ are not equal to $\epsilon_i$, but
they are related to them, 
\begin{equation}
H_L |f^{L}_{i\mathbf{k}_\perp}\rangle=\frac{1}{2}\mathrm{tanh}\frac{1}{2}
\epsilon_{Li}(\mathbf{k}_\perp)|f^L_{i\mathbf{k}_\perp}\rangle.\nonumber\\
\end{equation}

The Hamiltonian $H_L$ can be obtained from a Hamiltonian
in the whole space,
$H_{\mathrm{flat}}$ the flat-band Hamiltonian.  This Hamiltonian has
the same eigenfunctions as the true Hamiltonian but
has different eigenvalues, $-\frac{1}{2}$ for the occupied states and 
$+\frac{1}{2}$ for the empty ones.
$H_L$ is obtained from this by cutting off the right half of
the space;  cutting away the left half of the space leads
to a partner Hamiltonian $H_R$ 
(whose eigenfunctions $f_R$ generate the Schmidt states on the right).

The flat-band Hamiltonian has an unusual property--its
eigenstates can be reconstructed from the eigenstates of the two halves.
There is a correspondence $\mathcal{M}$ between the eigenfunctions
of $H_L$ and $H_R$ which reverses the sign of the ``energy". 
Using this correspondence, define
\begin{multline}
F_{i\mathbf{k}_\perp}(\mathbf{r})=\sqrt{\frac{1}{2}\mathrm{sech\ }
\frac{\epsilon_{Li}(\mathbf{k}_\perp)}{2}} \times\\
\left[ e^{-\frac{1}{4}\epsilon_{Li}(\mathbf{k}_\perp)}
f^L_{i\mathbf{k}_\perp}(\mathbf{r})+  
e^{\frac{1}{4}\epsilon_{Li}(\mathbf{k}_\perp)}
\mathcal{M}f^L_{i\mathbf{k}_\perp}(\mathbf{r})\right];
\label{Fff}
\end{multline}
then $F_{ik}$
is an eigenstate of $H_{\mathrm{flat}}$ 
with eigenvalue $-\frac{1}{2}$, and hence is occupied.
As $f^L_{i}$ varies over \emph{all} 
eigenstates of $H_L$, the function $F$
varies over a basis for the \emph{occupied}
states in the ground state\footnote{In one dimension, this
construction gives a basis of localized occupied states, somewhat
like Wannier orbitals.
Each of the $f^L$ and $f^R$ states, which have energies
in the gap, are confined to some layer near the surface. 
Their sum is an eigenstate of $H_{\mathrm{flat}}$. (Unlike
the Wannier basis, these states are all localized around
the same bond.)  The Hamiltonian $H_{\mathrm{flat}}$
has localized eigenfunctions, unlike most Hamiltonians, because
the dispersion relation is flat, and so the group velocity is equal to zero.}

When the system is inversion symmetric, $\mathcal{M}$ and $\mathcal{I}$
can be combined together to give the symmetry $\mathcal{I}_e$;
it is a transformation within the left half of the insulator,
defined by $\mathcal{I}_e=\mathcal{IM}$. Since $\mathcal{I}$
is a symmetry of the wave function, it preserves
$\epsilon$ while $\mathcal{M}$ reverses its sign.
Therefore, $\mathcal{I}_e$ acts as a particle-hole symmetry.

Now we return to the zero-``energy" states at TRIMs and their
$\mathcal{I}_e$-parities $\eta_{ei\bm{\kappa}_\perp}$.
Each state $f_{i\bm{\kappa}_\perp}$ 
extends, by Eq. (\ref{Fff}), to an occupied
state 
\begin{equation}
F_{i\bm{\kappa}_\perp}=\frac{1}{\sqrt{2}}
\left[(f^L_{i\bm{\kappa}_\perp}(\mathbf{r})+\eta_{ei\bm{\kappa}_\perp}
f^L_{i\bm{\kappa}_\perp}(-\mathbf{r})\right].\label{eq:openarms}
\end{equation}  This state is invariant
under $\mathcal{I}$, and the parity is $\eta_{ei\bm{\kappa}}$.

Let us determine the value of $\Delta N_e$ for a one-dimensional
system.  The result in higher dimensions follows since we can
use conservation of $\mathbf{k}_\perp$ to reduce the dimension to $1$.
Consider a circular chain with an even number of cells, $L$.
  Now, count
the number of even occupied states $W_e$ minus the number of odd
occupied states $W_o$,
using two different bases. $W_e-W_o$ is equal to $\mathrm{tr}\ \mathcal{I}$
so it is basis independent.  

One orthonormal basis will
be obtained by cutting the system along a diameter.
There will now be two cutting points $0$ and $\frac{L}{2}$.
  The zero-``energy" states give
parity eigenstates centered on each of the two cuts, according
to Eq. (\ref{eq:openarms}).  These
contribute $2\Delta N_e$ to $W_e-W_o$.  The remaining
states can be organized into inversion-related pairs,
$F_i(x),F_i(-x)$.  All these
states are mutually orthonormal because they correspond to different
 eigenvalues of $H_L,H_R$.
The inversion matrix $\mathcal{I}$
has only off-diagonal matrix elements between $F_i(\pm x)$, 
so they do not contribute to the trace.

On the other hand, instead of the localized wavefunctions,
we can use the extended
Bloch functions, $\psi_a(k_x)$.
The wave functions
at momentum $\pm k_x$ are exchanged, and the wave functions
at the TRIMs contribute $\Delta N(\pi)+\Delta N(0))$ to $W_e-W_o$.
Eq. (\ref{eq:half}) follows.

\begin{figure}
\includegraphics[width=.45\textwidth]{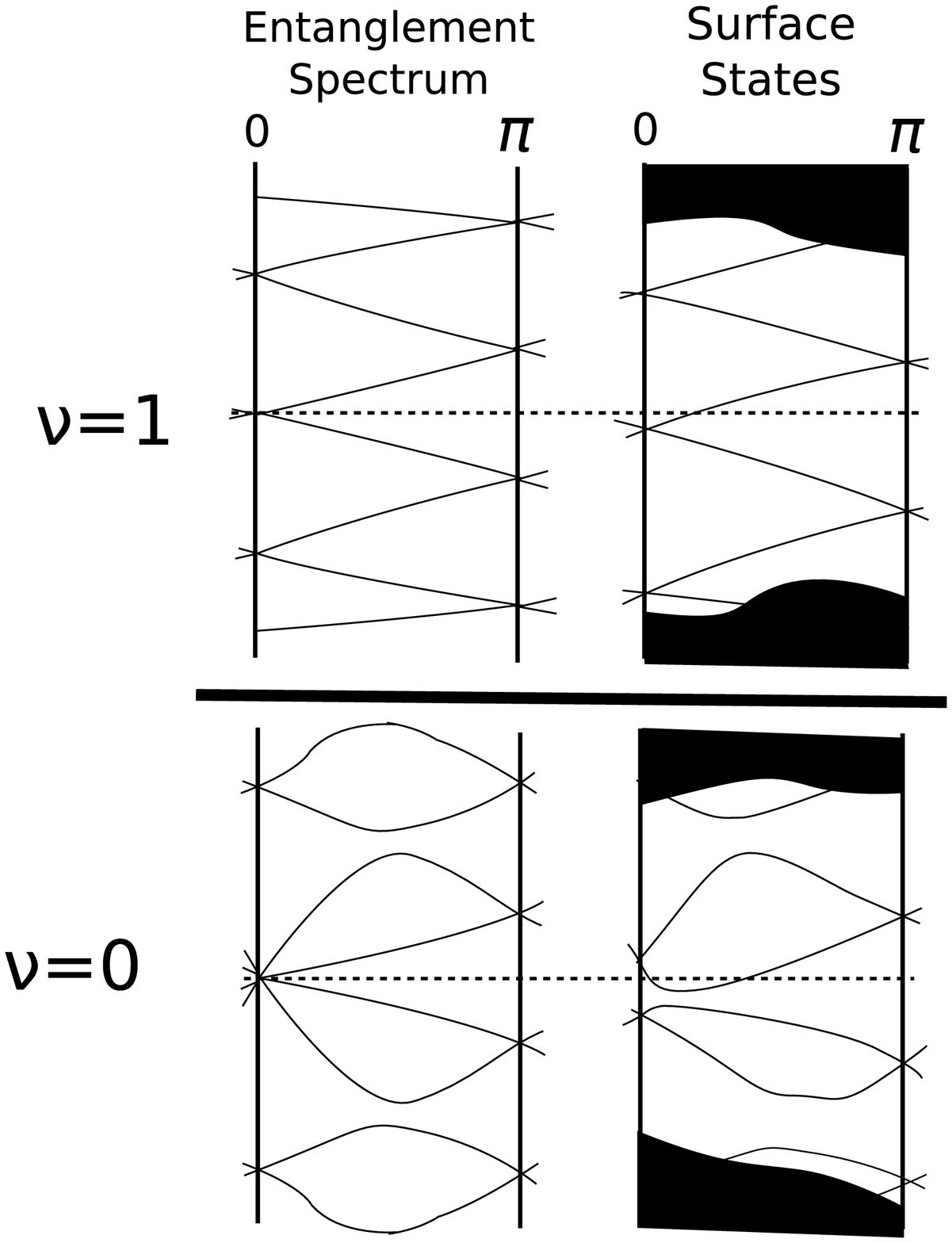}
\caption{\label{fig:slug}Determining the Quantum Spin Hall
index using the entanglement spectrum. The figure
compares spectra
of a nontrivial and a trivial system.  
The left spectrum for
each system is the entanglement spectrum, and the right
illustrates how the surface spectrum might look. In the entanglement
spectrum, inversion symmetry protects degeneracies at zero
energy at the
TRIMs, 
allowing one to determine the index.   But the two sets of
spectra can be deformed into one another (the difference
is probably more drastic than illustrated).
Because time reversal
symmetry produces Kramers degeneracies at the TRIMs at all energies,
the parity of the number of modes crossing the Fermi energy does not
change.}
\end{figure}

We can now count the edge states of a two-dimensional insulator 
that has both time reversal and inversion symmetry, to obtain
the formulae from Ref. \onlinecite{TI}
for the indices of topological insulators.  We will focus
on the two dimensional quantum spin-Hall index, since the three
dimensional indices are defined in terms of it. 
The quantum spin-Hall index $\nu$
is the number (modulo 2) of edge modes betweeen $0$ and $\pi$. 
Finding surface states
in the entanglement spectrum is easy because of the
extra particle-hole symmetry (see Fig. \ref{fig:slug}); these states remain
when the spectrum is deformed into the physical spectrum, by the
standard arguments. 

Consider the entanglement spectrum created by dividing the
system at $y=0$.
The index is the parity of the number of dispersion
curves crossing
a line $\epsilon=\mathrm{const.}$ between $k_x=0$ and $\pi$.  
Consider in particular the axis $\epsilon=0$.
Strictly between
$0$ and $\pi$ the axis crosses an even number
of modes: the crossings come in pairs because
the spectrum is symmetric under $\mathcal{T}\mathcal{I}_e$,
which just flips the sign of $\epsilon$.
(Generically, these states will just mix and move off the axis.)

Therefore only the modes at the ends of the interval are important.
We may assume that all the modes at either of these TRIMs have
the same parity, because otherwise the states whose
parity is in the minority may combine with the states
in the majority and become gapped. Then by Eq. (\ref{eq:half})
the number of modes at, e.g., $k_x=0$ is 
\begin{equation}
|\Delta N_e(0)|=|n-n_o(0,0)-n_o(0,\pi)|.
\label{eq:SE}
\end{equation}
Since these modes are at the extremes of the interval from
$0$ to $\pi$, they only qualify as half-modes.  One
way to see this is to look at a line slightly above the axis.
This line
crosses half of the modes emanating from each
TRIM, so the number of crossings, mod. 2,
is $\nu\equiv\frac{1}{2}\sum_{k_x=0}^\pi|n-n_o(k_x,0)-n_o(k_x,\pi)|$. This
is congruent to $\frac{1}{2}\sum_{\bm{\kappa}}n_o(\bm{\kappa})$, summed
over all four TRIMs.

When the flat band Hamiltonian is deformed into the true Hamiltonian, 
$\nu$ remains the same even though
the surface states no longer have particle-hole symmetry.
 The energy curves form continuous loops or zigzags (see figure)
because of the double-degeneracies protected by Kramers' theorem.

\begin{figure}
\includegraphics[width=.45\textwidth]{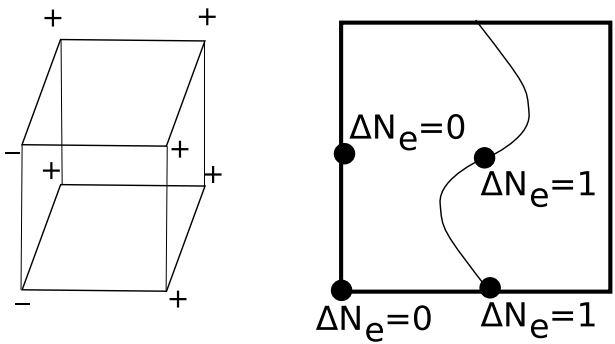}
\caption{\label{fig:lawnmower} A model
with a single filled band that
has an odd quantum Hall effect parallel to $\mathbf{R}_z$. 
a) The parities at the TRIMs.
b) The entanglement states on the $xz$-face of the Brillouin zone,
determined using $\Delta N_e(\bm{\kappa}_\perp)=\frac{1}{2}(\Delta N(\bm{\kappa_1}+\Delta N(\bm{\kappa}_2))$.
}
\end{figure}

We can use a similar approach to understand
the results for the polarization and Hall coefficient.
These effects 
may
be determined by sketching the arcs of the entanglement Fermi
surface rather than the real one. 
While each zero energy state at a TRIM
gives at least a Fermi point, not all of these points extend
to Fermi arcs.  It is certain that the \emph{parity} of the
number of Fermi arcs is the same as $\Delta N_e$'s parity however.
(This is proved in appendix \ref{app:fishbird}.)

Fig. \ref{fig:lawnmower} shows how the modes might look for a
set of
parities that corresponds to an odd quantum Hall conductance.
Consider the $xz$ surface of this insulator.
According to Eq. (\ref{eq:half}) there must be one arc (or
an odd number) passing through
$(\pi,0)$ and $(\pi,\pi)$, and an
even number through the other two TRIMs.
Hence as one travels along the $x$-direction through
the Brillouin zone, one crosses
an odd number of Fermi arcs. (Any extra arcs that do not pass
through TRIMs come in pairs by inversion symmetry.)  
Since the number of arcs (counted
with a sign depending on the sign of the group velocity)
is the $z$-component of the Hall conductivity, $\tilde{G}_z/(2\pi)$ is odd.

\begin{figure}
\includegraphics[width=.45\textwidth]{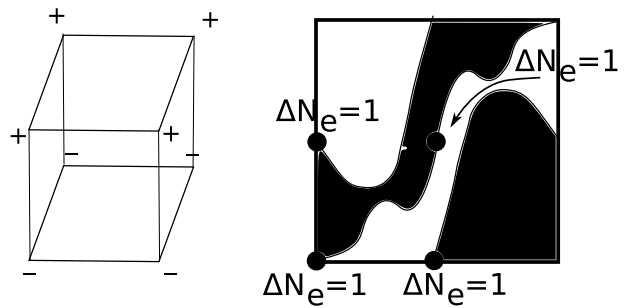}
\caption{\label{fig:polarization}
Determining the polarization from the entanglement spectrum.
a) Parities for an insulator with one filled band, with a half-filled
Fermi sea in the entanglement spectrum. 
b)Possible entanglement Fermi arcs for a cut parallel to the $xz$ plane.
The arcs surround half the surface
Brillouin zone. If there
are no nuclei on $y=0$, this crystal has a half-integer
polarization in the $y$-direction.
}
\end{figure}

Now consider the parities in Fig. \ref{fig:polarization}.  One possible
choice of modes is illustrated.  If these modes are not chiral,
then the states below energy $\epsilon=0$
cover half the Brillouin zone (exactly half
because of the symmetry). This will imply that $P^{y}=\frac{e}{2}$ if
there are no nuclei on $y=0$.

The polarization is defined
as the surface charge once the surface bands 
have been emptied.  (See appendix \ref{app:polarization}.) This
definition may be applied to the entanglement spectrum.
Cut the system with a plane $y=0$ through a point of inversion symmetry.
An important property of the entanglement spectrum is
that every term in the Schmidt decomposition has a net
charge of zero in the vicinity of the cutting surface.

Of all the terms in the Schmidt decomposition, consider
the one with the largest coefficient, $|G\rangle_L|G\rangle_R$,
where $|G\rangle_L$ and $|G\rangle_R$ are the
ground states of $H_L$ and $H_R$ respectively, obtained by
filling all the negative energy states.  (It does not
matter which of the states with exactly zero
energy are filled.)
 These states are mirror images of each other, so
they both have the same surface charge $Q_y$. The net
charge is $2Q_y$.  However, the Schmidt state
$|G\rangle_L|G\rangle_R$, obtained by a hypothetical
measurement of the two halves of the system, has
the same charge as the ground state.  Each electron is pushed
over to the side of the $y=0$ plane where it is more likely
to be, but no electrons appear or disappear. Thus, 
 $Q_y=0$.

Now the ground
state of $H_L$ has a partially filled band.  When this 
band (which covers half the
Brillouin zone) is emptied, a charge of $-\frac{e}{2}$ per unit
cell on the surface remains.
Hence the polarization $P^{y}\equiv \frac{e}{2}\ (\mathrm{mod\ }e)$. 

To compare this result to Eq. (\ref{eq:polarization}), we need
to take nuclei into account.  Note that $P_e^{y}=0$ for
the given parities. By neutrality, there
must be a single charge $-e$ nucleus per unit
cell, which contributes 
$P_{n}^{y}=-\frac{e}{2}$.
Appendix \ref{app:polarization} gives more details.

\emph{Conclusions}
This study of inversion symmetric insulators displays various
interconnections between quantized electrical properties and
the inversion parities at TRIMs.  In particular, it provides
a simple criterion for determining when a crystalline compound
has a magnetoelectric response equal to $\pi$.  The magnetoelectric
effect could be seen in a system with broken time-reversal symmetry,
even if the spin-orbit coupling is small.

The properties of inversion symmetric insulators can be derived
in a simple way with the help of the entanglement spectrum.

Future work could extend these results to other interesting lattice symmetries,
as well as the classification of superconductors with crystal symmetries.
Finally, 
understanding the stability of the parity invariants to
interactions will be an interesting subject for future study.

\emph{Acknowledgments}
This research
was supported by NSF-DMR-0645691 and NSF-DMR-0804413.
We thank Andrew Essin and Joel Moore
for conversations about topological responses and Sergey Savrasov
and Xiangang Wan for collaborating on a related project, thus helping
to shape this article as well. Parallel work by T. Hughes, E.V. Prodan and
B. A. Bernevig considers many of the same issues
about inversion-symmetric insulators\cite{Hughesentanglement}.

%
%


\appendix

\section{\label{app:americancheese}Classification using the topology of Hamiltonians}


In this section we will show how to use systematic methods from topology
to classify the insulators with inversion symmetry. This method
was also used to find the phases
of systems with one of the Altland and Zirnbauer symmetry
groups\cite{ten,eleven}.

In general we write a band insulator as $H(\vk)$, which we think of as a vector bundle on the Brillouin zone.
The eigenvectors of $H(\vk)$ with energies below the chemical potential are the 
occupied states, defining a vector subspace of the entire Hilbert space at 
$\vk$. For inversion-symmetric insuators, there is an additional constraint 
on the Hamiltonian: $\kRev^0[H(\vk)] \equiv I_0 H(\vk) I_0 = H(-\vk)$, this 
relates the Hamiltonians at $\vk$ and $-\vk$.

The idea behind our method of classification, similar to the one used for time-reversal invariant topological insulators, is to only look at half the Brillouin zone.

For a Hamiltonian $H$ in $d$\hyp{}dimensions, one can construct a path $f(t) = H(k_x=\pi t, k_2, k_3, \dots)$ in the space of $(d-1)$\hyp{}dimensional Hamiltonians, where the endpoints ($t = 0,1$) are inversion symmetric.
Thus the classification of $d$\hyp{}dimensional Hamiltonians is equivalent to the classification of paths in $\Path(\Hcal_{d-1}; \Ical_{d-1}, \Ical_{d-1})$, where $\Hcal_d$ is the set of general $d$\hyp{}dimensional Hamiltonians and $\Ical_d \subset \Hcal_d$ is the subset which is inversion-symmetric.
The symbol $\Path(X;A,B)$ is defined to be:
\begin{align}
	\Path(X; A,B) & \equiv \textrm{Set of paths in $X$ from $A$ to $B$}
		\\	&	= \{ f:[0,1] \rightarrow X | f(0) \in A, f(1) \in B \} ,			\notag
	\label{eq:PathSpaceDef}
\end{align}
where $A,B \subset X$.
We want to divide up $\Ical_d$ into sets such that paths from different sets are not homotopic to each other.
Heuristically, the classes of paths of $\Path(X;A,B)$ are given by the number of components of $A,B$ which determines the set of possible endpoints, and the loop sturcture of $X$ which determines the number of ways to travel from $A$ to $B$.
The classes of insulators in $d+1$\hyp{}dimensions is very roughly given by:
\begin{align}
	&	\textrm{Components of $\Ical_{d+1}$}				\notag
	\\	&\quad
		\sim \frac{\textrm{Loops in $\Hcal_d$}}{\textrm{Loops in $\Ical_d$}}
		\times (\textrm{Components of $\Ical_d$})^2 .
\end{align}
This idea is made precise using algebraic topology\cite{algebraictopology}, and is captured by the exact sequence \eqref{eq:ExactSeq_pi1XA}.


There is an addition structure in the classification of insulators\cite{algebraictopology}, which comes from the fact that one can combine two insulators together using direct sums ``$\oplus$''.
To simplify the classification, it is useful to also have a subtraction ``$\ominus$'' operation between insulators.  This would give the topological invariants (\textit{e.g.} $\ZZ$) a group structure.

The subtraction procedure is realized by considering an ordered pair of bands $(H_1,H_2)$, which represents the ``difference'' of the two Hamiltonians.
Addition by $H'$ is given by $(H_1 \oplus H',H_2)$ and subtraction is given by $(H_1, H_2 \oplus H')$.
Imposing the equivalence relation $(H_1 \oplus H',H_2 \oplus H') \sim (H_1,H_2)$ makes the addition and subtraction processes cancel each other.
Physically, we are interested in classifying difference of two topological insulators -- this is analogous to studying domain walls
whose properties are determined only by
the difference in topological invariants.
With this interpretation, it is possible to talk about a negative number of 
filled bands (whenever $H_2$ has more bands than $H_1$).

The construction above, called the \emph{Grothendieck group}, 
yields two direct results.
First, two insulators $H_1,H_2$ are deformable to one another when the toplogical invariants of the band structure $(H_1,H_2)$ are all trivial.
Second, when the latter insulator is the vacuum, the topological invariants 
of $(H,\textrm{vac})$ are simply the bulk band invariants of $H$.


\subsection{Hamiltonians, Classifying Spaces and Homotopy Groups}

Consider an $N \times N$ matrix $H$ with $n$ occupied states and $N-n$ empty states.
Setting the chemical potential to be zero, $H$ is a matrix with $n$ negative eigenvalues and $N-n$ positive eigenvalues.
In the topological classification of insulators, the energies are irrelevant so long as we can distinguish between occupied and unoccupied states, and hence we can deform the energies (eigenvalues) of all valence bands to $-1$ and the energies of conduction bands to $+1$.
We can also assume there are an infinite number of conduction bands, and so we let $N \rightarrow \infty$.
$\Hcal_0$ is the set of such 0D Hamiltonians, and can be separated into discrete components based on the number of filled bands (which may be negative).
The space $\Hcal_0$ is homeomorphic to $\ZZ \times BU$, where $BU$ is the classifying space of the unitary groups.
At the TRIM, the Hamiltonian is inversion-symmetric and commutes with the operator $I_0$.  The Hilbert space is divided into an even subspace
and an odd subspace, based on the inversion eigenvalues.
Hence the set of inversion-symmetric (0D) Hamiltonians $\Ical_0$ is homeomorphic to $\Hcal_0 \times \Hcal_0$.

Finally, we introduce the ``vacuum'' $v_0 \in \Ical_0 \subset \Hcal_0$, which is a Hamiltonian with no filled bands.
$v_0$ is a useful object in that it allows us to compare any Hamiltonian to it, and also acts as the basepoint when we compute the homotopy groups of $\Hcal_0, \Ical_0$.

Given a topological space $X$ and a basepoint within the space $x_0$, the \emph{homotopy group} $\pi_s(X)$ is the set of
equivalence classes of maps $f: (S^s, b_0) \rightarrow (X, x_0)$, where the basepoint $b_0 \in S^s$ and $f(b_0) = x_0$.
For example, $\pi_0(X)$ gives the number of connected components of $X$, $\pi_1(X)$ tells us which loops in $X$ are equivalent and which loops are contractible.

The group strucutre of $\pi_s(X)$ is given by concatenation of maps.
However, the group structure of the Hamlitonians has already been defined based on direct sums.
Fortunately, the group composition defined based on the two methods (concatenation / direct sums) are compatible.


The homotopy groups of $\Hcal_d$ are known:
$\pi_0(\Hcal_0) = \ZZ$ because 0-dimensional Hamiltonians are classified by 
the number of filled bands $n$.
$\pi_1(\Hcal_0) = 0$ means that the loops in $\Hcal_0$ are all contractible.
$\pi_2(\Hcal_0) = \ZZ$, because maps of the sphere are classified by the 
first Chern class (or the Chern number).  This invariant give rise to the integer quantum Hall effect.
For higher dimension, $\pi_s(\Hcal_0)$ is $0$ when $s$ is an odd integer, and $\ZZ$ when $s$ is an even integer, corresponding to the $(s/2)$\textsuperscript{th} Chern class.  In this section, Chern numbers are taken to be integers rather
than multiples of $2\pi$.

The homotopy groups of $\Ical_0$ are simply the squares of the homotopy groups of $\Hcal_0$.
In particular, the set of components $\pi_0(\Ical_0) = \ZZ^2$ is labeled by two integers: $(n, \alpha)$, where $n$ is the total number of valence ``bands" 
and $\alpha = n_o$ is the number of states which have odd inversion parity.

\subsection{Relative Homotopy Groups and Exact Sequences}

\begin{figure}
\includegraphics[width=.25\textwidth]{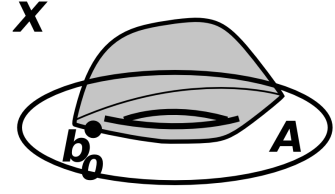}
\caption{Relative homotopy groups.  The figure represents an element of
$\pi_2(X,A,x_0)$ where $X$ is $\mathbb{R}^3$; $A$ is the torus. 
\label{fig:relatives}}
\end{figure}

The homotopy groups $\pi_s(X)$ classifies components, loops, and maps from higher dimensional spheres to the space $X$.
The \emph{relative homotopy groups} $\pi_s(X,A)$ classify maps from paths, disks, \textit{etc.} where the boundary must lie in some subspace (see Fig. \ref{fig:relatives}). This is how topologists define ``winding numbers" of
open arcs, which were discussed
in Sec. \ref{sec:class}. Relative homotopy groups were
applied much earlier by Refs. \cite{volovik} to an interesting problem
within physics: classifying defects of ordered phases when
the defects are stuck to the surface.

Given a space $X$ a subspace $A \subset X$, and a basepoint $x_0 \in A$, the relative homotopy group $\pi_s(X,A)$ is the equivalence classes of maps $(D^s, \partial D^s, b_0) \rightarrow (X, A, x_0)$.
The boundary of the disk $\partial D^s = S^{s-1}$ must map to $A$, and the basepoint $b_0 \in \partial D^s$ maps to $x_0$.
The relative homotopy groups can be computed via the exact sequence
\begin{align}
	\pi_s(A)
		\xrightarrow{i_\ast} \pi_s(X)
		\xrightarrow{j_\ast} \pi_s(X,A)
		\xrightarrow{\partial} \pi_{s-1}(A)
		\xrightarrow{i_\ast} \pi_{s-1}(X) .
	\label{eq:ExactSeq_pisXA}
\end{align}
An \emph{exact sequence} is a sequence of groups along with maps defined from one group to the next, each map preserving the group operations.
In an exact sequence, the image of every map is the same as the kernel of the subsequent map.

In the one-dimensional case, the relative homotopy group $\pi_1(X,A)$ describes the set of possible paths in $X$ from $x_0$ to $A$ up to homotopy,
that is, the classes of paths $\Path(X;x_0,A)$.
The exact sequence becomes:
\begin{align}
	\pi_1(A)
		\xrightarrow{i_{(1)}} \pi_1(X)
		\xrightarrow{j_{(1)}} \pi_1(X,A)
		\xrightarrow{\partial} \pi_0(A)
		\xrightarrow{i_{(0)}} \pi_0(X) .
	\label{eq:ExactSeq_pi1XA}
\end{align}

In the exact sequence above, the maps are defined as follows.
\begin{itemize}
\item	$i: A \rightarrow X$ is the inclusion map which takes every point from $A$ to a point in $X$.
The induced maps $i_{(s)}: \pi_s(A) \rightarrow \pi_s(X)$ and takes components/loops from one space to the other.
\item	All the loops in $X$ start and end at $x_0$, and so they are also clearly paths in $\Path(X;x_0,A)$ seeing $x_0 \in A$.
$j_{(1)}$ is the map that takes the equivalence classes of loops $\pi_1(X)$ to the equivalence classes of paths $\pi_1(X,A)$.
\item	$\partial: \pi_1(X,A) \rightarrow \pi_0(A)$ is a map that takes a path and selects its endpoint to give a component of $A$.
$\partial$ is called the boundary map, it takes a ``1D object'' to give a ``0D object.''
\end{itemize}
It appears that the maps $i$ and $j$ ``do nothing'' to the objects (points, loops) they act on.
However, each map gives the loop/path more freedom to move around.  For example, $j_{(1)}$ takes a loop to a path where the endpoints no longer have to be the same.

\begin{figure}[tb]
	\includegraphics[width=0.23\textwidth]{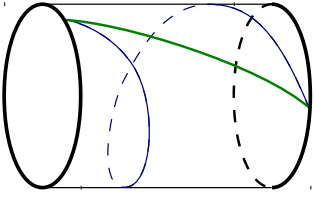}
	\caption{%
Understanding the exact sequence.
Paths in $\pi_1(X,A)$ can be enumerated by taking one path
for each inequivalent end-point and adding loops from $\pi_1(X)$
to the paths. This description is sometimes redundant.
In this figure, $X$ is the cylinder and $A$ is the two circles at its ends. 
A path $p$ and a path $p'=p+l$ obtained from it are shown where $l$
winds around the cylinder.
The two paths are equivalent because they can be deformed
into one another by bringing the right end-point around $A$.
This happens because $l$ can be smooshed into $A$.
	\label{fig:pi1Xunwound}}
\end{figure}

The exact sequence captures the idea that the paths in $\Path(X; x_0, A)$ 
can be classified once one knows the properties of $X$ and $A$,
based on their end-points
and how they wind.
\begin{enumerate}
\item
First, we pick the endpoint $x \in A$ of the path $p$.
$x$ must be connected to $x_0$.
and this is captured by the statement $\ker(i_{(0)}) = \img(\partial)$.
The explanation is as follows.
The endpoint of the path is given by the $\partial$ map: $x = \partial p$.
The image of $\partial$ is thus the set of elements of $\pi_0(A)$
that are connected to $x_0$ within $X$. On the other hand, these are exactly
the components which become equal to $0$ in $\pi_0(X)$, when $A$ is enlarged.
Hence $\ker(i_{(0)}) = \img(\partial)$.


\item
Given a choice of a path $p$ from $x_0$ to $x$, we can construct all the other paths between the points.
We can create any other path $p'$ by concatenating a loop $l \in \pi_1(X)$ at the beginning of $p$.
\item
However, the paths $p$ and $p'$ are only different (\textit{i.e.} not homotopic to each other) only if the loop $l$ cannot be unwound within $A$, this is to say that $p \sim p'$ if $l$ is homotopic to a loop in $A$. See 
Fig. \ref{fig:pi1Xunwound}.  This idea is captured by the exactness at $\pi_1(X)$.  Hence we think of $l$ belonging to 
the quotient $\frac{\pi_1(X)}{i_{(1)}\pi_1(A)}$, and this group is called
 the \emph{cokernel} of the map $i_{(1)}$.
\end{enumerate}

We see that any path may be constructed by its endpoint $x$ and a loop $l$.
\begin{align}
	p = j(l) + \partial^{-1}x
	.	\label{eq:PathLoopEndpoint}
\end{align}
The inverse boundary operator $\partial^{-1}$ is not unique, but that does not affect the structure of the group $\pi_1(X,A)$ for the cases we are considering.
The equivalence classes of $x$ form $\ker(i_{(0)})$, while the equivalence classes of $l$ make up $\coker(i_{(1)})$.
The relative homotopy group is a semidirect product
\begin{align}
	\pi_1(X,A) = \coker(i_{(1)}) \rtimes \ker(i_{(0)}) .
\end{align}
What this means is that $\coker(i_{(1)})$ is a normal subgroup of $\pi_1(X,A)$ and $\ker(i_{(0)})$ is the quotient of the two.
For the purpose of classifying inversion-symmetric insulators, we can treat the semidirect product as simply a product.



\subsection{One Dimension}

In this section we examine the classification of 1D inversion-symmetric Hamiltonians $H(k)$.
Let $\Ical_1$ be the set of maps $H: S^1 \rightarrow \Hcal_0$ such that $\kRev^0[H(k)] = H(-k)$.
$\Ical_1$ is homeomorphic to the set of paths in $\Hcal_0$ that start and end in $\Ical_0$ (\textit{i.e.} $\Ical_1 \approx \Path(\Hcal_0; \Ical_0, \Ical_0)$), and we seek to classify all such paths - to compute $\pi_0(\Ical_1)$.

For a 1D band structure $H(k)$, we can `factor' out $H(0)$ by letting $H'(k) = H(k) \ominus H(0)$ such that $H'(0) = v_0$.
$H(0)$ is an element of $\Hcal_0$ while $H'(k)$ is an element of $\tI_1$, where
$\tI_1$ is the subset of $\Ical_1$ with a fixed basepoint ($H'(0) = v_0$).
The decomposition $H(k) = H(0) \oplus H'(k)$ can be expressed as
\begin{align}
	\Ical_1 = \Ical_0 \times \tI_1 .
\end{align}
Hence the classification of $\Ical_1$ can be broken up in to two parts, the classification of $\Ical_0$ and that of $\tI_1$.
The former is understood as $\pi_0(\Ical_0) = \ZZ^2$.

Notice that $\tI_1$ is homeomorphic with the class of paths $\Path(\Hcal_0; v_0, \Ical_0)$, whose
components are described by the relative homotopy group $\pi_1(\Hcal_0, \Ical_0)$. These correspond
to the possible phases of band structures in $\tI_1$.
The relative homotopy group can be computed by the exact sequence \eqref{eq:ExactSeq_pi1XA}.
Using the fact that $\pi_0(\tI_1) = \pi_1(\Hcal_0, \Ical_0)$, 
\begin{align}
	\ColTwo{\pi_1(\Ical_0)}{0}
		\xrightarrow{i^1_\ast} \ColTwo{\pi_1(\Hcal_0)}{0}
		\xrightarrow{j^0_\ast} \pi_0(\tI_1)
		\xrightarrow{\partial} \ColTwo{\pi_0(\Ical_0)}{\ZZ^2}
		\xrightarrow{i^0_\ast} \ColTwo{\pi_0(\Hcal_0)}{\ZZ} .
	\label{eq:ExactSeqI1}
\end{align}
Since $\pi_1(\Hcal_0) = 0$, we can ignore the left side of the exact sequence (cokernel of $i_{(1)}$ is zero).
The integer $n \in \pi_0(\Hcal_0) = \ZZ$ tells us how many filled bands there are, and $(n_o, n_e) \in \pi_0(\Ical_0) = \ZZ^2$ are the number of even\hyp{}parity and odd\hyp{}parity bands.
The map $i_{(0)}$ is given by $n = n_o + n_e$, and so the kernel of the map is the subset where $n_o = -n_e$, isomorphic to 
$\ZZ$.~\footnote{The Hamiltonian we are classifying ($H'(k) =H(k) \ominus 
H(0)$)
 is one of the generalized Hamiltonians defined above; that is
why a number of bands $n_e$ or $n_o$ can be negative.}
It is clear that $\pi_0(\tI_1)$ is isomorphic to $\ker(i_{(0)}) = \ZZ$.
This is to say that the set of paths $\Path(\Hcal_0; v_0, \Ical_0)$ are solely classified by the endpoint.
Since $\Ical_1 = \Ical_0 \times \tI_1$, we have
\begin{align}
	\pi_0(\Ical_1) & = \pi_0(\Ical_0) \times \pi_0(\tI_1)			\notag
		\\	& = \pi_0(\Ical_0) \times \pi_1(\Hcal_0, \Ical_0)			\notag
		\\	& = \ZZ^2 \times \ZZ .
\end{align}
The invariant $\pi_0(\Ical_0) = \ZZ^2$ corresponds to the number of total bands and odd\hyp{}parity states at $k=0$: $n, n_o(0)$.
The invariant $\pi_0(\tI_0) = \ZZ$ corresponds to the difference in number of odd bands at $k=\pi$ and $k=0$: 
$\alpha_x=n_o(\pi) - n_o(0)$.
Hence the parities at the two TRIMs completely classify all 1D inversion-symmetric Hamiltonians.

A \emph{generator} of a group is an element which gives the entire group by group addition and subtraction, 
for example $1$ is a generator of $\ZZ$.
In our case, the generators are Hamiltonians which forms a basis for all Hamiltonians, up to homotopy.

The generators of $\pi_0(\Ical_0) = \ZZ^2$ are two Hamiltonians $H_n^0$ and $H_\alpha^0$ which adds one even-parity and odd-parity band to the system, respectively.
Explicitly:
\begin{align}
	H_n^0 & = \begin{bmatrix} -1 \end{bmatrix}_{(+)}
	.	\label{eq:Generatorn}	\\
	H_\alpha^0 & = \begin{bmatrix} -1 \end{bmatrix}_{(-)} \ominus H_n^0
	.	\label{eq:GeneratorAlpha0}
\end{align}
where the subscript $(\pm)$ labels the inversion eigenvalue(s) of the orbital(s).
The first expression $H_n^0$ adds an inert band  
to increase $n$; the second expression $H_\alpha^0$ adds an odd-parity band but subtracts an even-parity one to increase $n_o$ while maintaining $n$.
The generator for $\pi_0(\tI_1) = \ZZ$ is:
\begin{align}
	H_\alpha^1(k) = \begin{bmatrix} -\cos k & \sin k \\ \sin k & \cos k \end{bmatrix}_{(+-)} \ominus H_n^0
	.	\label{eq:GeneratorAlpha1}
\end{align}
The subscript $(+-)$ specifies the inversion operator $I_0 = \SmallMxTwo{+1 &}{& -1}$ for this Hamiltonian.
When $k = 0$, the matrix becomes $\SmallMxTwo{-1 & }{ & 1} \!\phd_{(+-)}$ and the occupied band is even under inversion.  Similarly, the matrix at $k = \pi$ is $\SmallMxTwo{1 & }{ & -1} \!\phd_{(+-)}$ and there the occupied band is odd.  Hence $\alpha_x = n_o(\pi) - n_0(0) = 1 - 0 = 1$ and $H_\alpha^1(k)$ is a generator of $\pi_0(\tI_1)$.

Therefore, any 1D inversion-symmetric Hamiltonian is homotopic to
\begin{align}
	H(k) = n H_n^0 \oplus \alpha H_\alpha^0 \oplus \alpha_x H_\alpha^1(k) .
\end{align}

\subsection{Two Dimensions}

We apply the same ideas used in 1D to classify inversion-symmetric insulators in 2D.
The inversion-symmetric Hamiltonians $(k_x,k_y)$ in 2D satisfy: $\kRev^0 H(k_x,k_y) = H(-k_x,-k_y)$.
The set of 2D inversion symmetric Hamiltonians ($\Ical_2$) is equivalent to $\Path(\Hcal_1; \Ical_1, \Ical_1)$, where $\Hcal_1$ is the set of 1D band structures (loops in $\Hcal_0$).

Just as in the 1D case where we decompose $H(k)$ into a 0D and 1D object: 
$H(k) = H(0) \oplus H'(k)$, we decompose the 2D Hamiltonian into 0D, 1D and 2D components.
Let 
\begin{subequations}
\begin{align}
	H'(k_x,k_y) = H(k_x,k_y) \ominus H(0,0) ,
\end{align}
so that $H'(0,0) = v_0$.
Now we define
\begin{align}
	H''(k_x,k_y) = H'(k_x,k_y) \ominus H'(0,k_y) \ominus H'(k_x,0) ,
\end{align}
\end{subequations}
so that
\begin{align}
	H''(0,k_y) = H''(k_x,0) = v_0 .
	\label{eq:tI_2Def}
\end{align}
$H(0,0)$ is an element of $\Ical_0$, and $H'(0,k_y)$ and $H'(k_x,0)$ are elements of $\tI_1$.
We define $\tI_2$ to be the set of inversion-symmetric Hamiltonians satisfying \eqref{eq:tI_2Def}.
With this procedure, we have decomposed $\Ical_2$ as
\begin{align}
	\Ical_2 = \tI_0 \times \tI_1^2 \times \tI_2 ,
		\label{eq:Ical2Decomp}
\end{align}
where $\tI_0 = \Ical_0$.
Explicitly, the decomposition is:
\begin{align}
	& H(k_x,k_y)			\notag
	\\	& = \underbrace{ H(0,0) }_{\tI_0}
		\oplus \underbrace{ H'(0,k_y) \oplus H'(k_x,0) }_{\tI_1 \times \tI_1}
		\oplus \underbrace{ H''(k_x,k_y) }_{\tI_2}
\end{align}
Due to \eqref{eq:tI_2Def}, we can think of Hamiltonians in $\tI_2$ as maps from the sphere (rather than the torus) to $\Hcal_0$.

We now analyze the properties of Hamiltonians in $\tI_2$.
For each fixed $k_x$, the Hamiltonian $H''(k_y)|_{k_x}$ is a map from the 1D Brillouin zone ($S^1$) to $\tH_0$ (where we've also defined $\tH_0 = \Hcal_0$).
Denote the set of such maps as $\tH_1$, the set of loops in $\tH_0$ with basepoint $v_0$: $\tH_1 = \Path(\tH_0; v_0, v_0)$.
$\tH_1$ is called the \emph{loop space} of $\tH_1$, and the notation used in literature is $\tH_1 = \Omega \tH_0$.

The Hamiltonian at $k_x = \pi$ is inversion-symmetric, and so $H''(k_y)|_{k_x = \pi} \in \tI_1$.
At $k_x = 0$, the line $H''(k_y)|_{k_x} = v_0$ is a constant map - which we call $v_1$ (a line of $v_0$).
Clearly $v_1$ is an element of $\tI_1$, and acts as the basepoint when we compute the homotopy groups of $\tI_1, \tH_1$.

Having defined the spaces $\tI_2, \tH_1$ and basepoint $v_1$, we can see that $\tI_2$ is homeomorphic to the set of paths in $\tH_1$ with endpoints at $v_1$ and $\tI_1$: $\tI_2 \approx \Path(\tH_1; v_1, \tI_1)$.
The exact sequence which gives the equivalence classes of such paths is
\begin{align}
	\pi_1(\tI_1)
		\xrightarrow{i^1_\ast} \pi_1(\tH_1)
		\xrightarrow{j^1_\ast} \pi_0(\tI_2)
		\xrightarrow{\partial} \pi_0(\tI_1)
		\xrightarrow{i^1_\ast} \pi_0(\tH_1) .
	\label{eq:ExactSeqI2}
\end{align}

Elements of $\pi_1(\tH_1)$ are two-dimensional Hamiltonians that equal $v_0$ along $k_x=0\sim2\pi$ and $k_y=0\sim2\pi$, which are equivalent to maps $S^2 \rightarrow \tH_0$.
Hence
\begin{align}
	\pi_1(\tH_1) & = \pi_2(\tH_0) = \ZZ	&&\textrm{(Chern number).}
\end{align}
The map $j^1$, which essentially maps general two-dimensional
Hamiltonians to inversion symmetric ones,
is defined by
\begin{align}
	& (j^1 H)(k_x,k_y)		\notag
	\\	&\quad
		= \begin{cases}
			H(2 k_x,k_y)	&	0 \leq k_x \leq \pi	\\
			\kRev^0 H(4\pi - 2k_x, 2\pi - k_y)	&	\pi \leq k_x < 2\pi
		\end{cases} .
	\label{eq:MapDef_j11}
\end{align}
It builds an inversion symmetric Hamiltonian out of two copies of a Hamiltonian with no special symmetries.
The map $\partial: \tI_2 \rightarrow \tI_1$ is defined by
\begin{align}
	[ \partial H ] (k) & = H(k_x=\pi, k_y=k)
\end{align}
which takes a 2D Hamiltonian and picks out the 1D Hamiltonian at $k_x = \pi$.



\begin{figure}[tb]
	\includegraphics[width=0.23\textwidth]{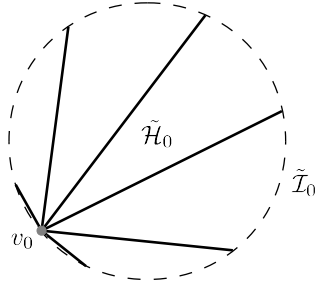}
	\caption{%
		Isomorphism between $\pi_2(\tH_0,\tI_0)$ and $\pi_1(\tI_1)$.
		An element in $\tI_1$ is a path from $v_0$ to an element of $\tI_0$ (dark black lines).
		An element of $\pi_1(\tI_1)$ is a family of such paths.
		The first path in the family must be $v_1$, a ``zero-length'' constant path at $v_0$ (gray dot).
		The last path of the family must also be $v_1$, and so the paths must shrink back to $v_0$ at the end, because that is the base point for the homotopy groups.
		This family of paths traces out a disk $D^2$ which maps to $\tH_0$, the endpoints of each path traces out a circle $S^1 = \partial D^2$ which maps to $\tI_0$ (dashed circle).
		The basepoint of the circle maps to $v_0$, and hence every element of $\pi_1(\tI_1)$ is also an element of $\pi_2(\tH_0,\tI_0)$ and vice-versa.
		This establishes the isomorphism $\pi_2(\tH_0,\tI_0) = \pi_1(\tI_1)$.
	}
	\label{fig:pi2H0I0_pi1I1}
\end{figure}
To solve the exact sequence earlier \eqref{eq:ExactSeqI2}, we need to compute $\pi_1(\tI_1)$,
which classifies loops of 1D inversion-symmetric insulators (not to be confused with 2D insulators).
As argued in Fig.~\ref{fig:pi2H0I0_pi1I1}, the group $\pi_1(\tI_1)$ is isomorphic to $\pi_2(\tH_0, \tI_0)$.
We can compute $\pi_1(\tI_1) = \pi_2(\tH_0, \tI_0)$ via the exact sequence \eqref{eq:ExactSeq_pisXA}.
\begin{align}
	\ColTwo{ \pi_2(\tI_0) }{\ZZ^2}
		\xrightarrow{i^0_\ast} \ColTwo{ \pi_2(\tH_0) }{\ZZ}
		\xrightarrow{j^0_\ast} \pi_1(\tI_1)
		\xrightarrow{\partial} \ColTwo{ \pi_1(\tI_0) }{0}
		\xrightarrow{i^0_\ast} \ColTwo{ \pi_1(\tH_0) }{0}
\end{align}
The first map $i^0_{(2)}: \pi_2(\tI_0) \rightarrow \pi_2(\tH_0)$ is given by $\beta = \beta_e + \beta_o$ (which means that the Chern numbers of the odd and even bands add to the total.)
The image of $i^0_{(2)}$ is all of $i^0_{(2)}$, so its cokernel is trivial.
Since on the right side the groups are also trivial ($\ker(i^0_{(1)}) = 0$), we have that $\pi_1(\tI_1) = 0$.



We now return to the exact sequence \eqref{eq:ExactSeqI2} to compute the $\pi_0(\tI_2)$.
\begin{align}
	\ColTwo{ \pi_1(\tI_1) }{0}
		\xrightarrow{i^1_\ast} \ColTwo{ \pi_0(\tH_2) }{\ZZ}
		\xrightarrow{j^1_\ast} \pi_0(\tI_2)
		\xrightarrow{\partial} \ColTwo{ \pi_0(\tI_1) }{\ZZ}
		\xrightarrow{i^1_\ast} \ColTwo{ \pi_0(\tH_1) }{0} .
\label{eq:2dexactseq}
\end{align}
The exact sequence yields $\pi_0(\tI_2) = \ZZ \times \ZZ$; the set of $H''(\vk)$ are classified by two integers $(\alpha_{xy}, \beta_{xy})$.
The first integer $\alpha_{xy}$ coming from the map $\partial$ gives $n_o(\pi,\pi)$; the second integer $\beta_{xy}$ is related to the Chern number of $H''$.
The basis of $\pi_0(\tilde{I}_2)$ are found by taking the image of the generator in $\pi_1(\tH_1)$ and one of the preimages 
of the generator element in $\pi_0(\tilde{I}_1)$, which we denote by $H_\beta^2(\vk)$ and $H_\alpha^2(\vk)$ respectively.
Any Hamiltonian in $\tI_2$, up to a homotopy, can be written as
\begin{align}
	H'' = \beta_{xy} H_\beta^2\oplus \alpha_{xy} H_\alpha^2 .
	\label{eq:H2alphabeta}
\end{align}
The generator $H_\beta^2(\vk) = i_{(1)}^1 H_{\textrm{chern}}(\vk)$ where $H_\textrm{chern}$ is the generator of $\pi_1(\tH_1) = \pi_2(\tH_0)$, \textit{i.e.} a 2D band insulator with Chern number $+1$.
$H_\alpha^2(\vk)$ is defined such that $(\partial H_\alpha^2)(k) = H_\alpha^1(\pi,k)$ is the 1D insulator \eqref{eq:GeneratorAlpha1}.





Explicitly, the generator
\begin{align}
	H_\alpha^2(\vk) & = \frac{1}{1 + x^2 + y^2} \begin{bmatrix}
			1 - x^2 - y^2 & 2(y + ix) \\
			2(y - ix) & x^2 + y^2 - 1
		\end{bmatrix}_{(+-)}	,			\notag
		\\	&\quad	\ominus H_n^0 ,
			\label{eq:GeneratorAlpha2}
\end{align}
where $x = \cot \frac{k_x}{2}, y = \cot \frac{k_y}{2}$.
When $x$ or $y = 0$, then $x^2 + y^2 \rightarrow \infty$ and $H_\alpha^2(\vk) = \SmallMxTwo{-1 &}{& 1} \!\phd_{(+-)}$ and $n_o(\vk) = 0$.
When $x = y = \pi$, it can be seen that $H_\alpha^2(\vk) = \SmallMxTwo{1 &}{& -1} \!\phd_{(+-)}$ and the filled band is odd under inversion.
Note that the Hamiltonian has Chern number $+1$. Since $\alpha_{xy}=n_o(\pi,\pi)$ (and the other
TRIMs have  $n_o=0$ because of the normalization), this
is related to the constraint $\tilde{G}\equiv \sum_{\bm{\kappa}} n_o(\bm{\kappa})$.

The other generator $H_\beta^2(\vk) = i_{(1)}^1 H_{\textrm{chern}}(\vk)$:
\begin{align}
	H_\beta^2(k_x, k_y) = H_\alpha^2(2k_x, k_y)
		.	\label{eq:GeneratorBeta2}
\end{align}
has Chern number $+1$ in each of the halves $0 \leq k_x \leq \pi$ and $\pi \leq k_x \leq 2\pi$.
Evidently, the Chern number of the entire Brillouin zone is given by
\begin{align}
	\tilde{G} & = \alpha_{xy} + 2\beta_{xy} ,
		\label{eq:ChernAlphaBeta}
\end{align}
since $H_\alpha^2, H_\beta^2$ has Chern number $1,2$ respectively.


The decomposition \eqref{eq:Ical2Decomp} gives us six $\ZZ$ invariants in 2D:
two from $\tI_0$, one from each of the two $\tI_1$, and two more from $\tI_2$.
The six invariants ($n, \alpha, \alpha_x, \alpha_y, \alpha_{xy}, \beta_{xy}$) are related to the properties
of the original Hamiltonian $H(k_x,k_y)$ as follows:
\begin{itemize}
	\item	$n$ gives the number of filled bands, generated by \eqref{eq:Generatorn}.
	\item	$\alpha = n_o(0,0)$ is the number of odd-parity states at $(k_x,k_y) = (0,0)$, generated by \eqref{eq:GeneratorAlpha0}.
	\item	$\alpha_x = n_o(\pi,0)-n_o(0,0)$ involves the difference between two parities, generated by \eqref{eq:GeneratorAlpha1} ($k \rightarrow k_x$).
	\item	$\alpha_y = n_o(0,\pi)-n_o(0,0)$ involves the difference between two parities, generated by \eqref{eq:GeneratorAlpha1} ($k \rightarrow k_y$).
	\item	$\alpha_{xy} = n_o(\pi,\pi) - n_o(\pi,0) - n_o(0,\pi) + n_o(0,0)$ involves the parities at all TRIMs, generated by \eqref{eq:GeneratorAlpha2}.
	\item	$\beta_{xy}$ relates to the Chern number: $\tilde{G} = 2\beta_{xy} + \alpha_{xy}$, generated by \eqref{eq:GeneratorBeta2}.
\end{itemize}

The rule for the Chern number's parity, Eq. (\ref{eq:narrowhall}), follows from the last constraint here.
\subsection{Going to Higher Dimensions}

In $d$\hyp{}dimensions, we want to calculate $\pi_0(\tI_d)$, the
set of components of the space of inversion-symmetric Hamiltonians.
To calculate this in a larger dimension, we will 
need to know $\pi_s(\tI_{d-s})$
in lower dimensions.

The relevant spaces are $\tH_d$ and $\tI_d$.
The general Hamiltonian space $\tH_d = \Omega \tH_{d-1}$ is the space of loops
of Hamiltonians in $\tH_{d-1}$, which become trivial on $k_x=0$ as well as on all
the other boundaries $k_i=0$ of the Brillouin zone.
Their homotopy groups are given by $\pi_s(\tH_d) = \pi_{s+d}(\tH_0)$.

The homotopy groups of the
subspace $\tI_d \subset \tH_d$ are harder to find.
This space
is homeomorphic to $\tI_d \approx \Path(\tH_{d-1}; v_{d-1}, \tI_{d-1})$,
since half of the Brillouin zone determines the Hamiltonian.
The homotopy groups of $\tI_d$ are therefore given by
the relative homotopy groups 
$\pi_s(\tI_d) \approx \pi_{s+1}(\tH_{d-1},\tI_{d-1})$.
Via the relative homotopy exact sequence \eqref{eq:ExactSeq_pisXA}, one can relate $\pi_{s+1}(\tH_{d-1},\tI_{d-1})$ to $\pi_s,\pi_{s+1}(\tH_{d-1})$ and $\pi_s,\pi_{s+1}(\tI_{d-1})$.
The homotopy structure of $\tI_d$ depends on that of $\tI_{d-1}$. Iterating this process 
reduces the problem to that of the basic objects $\tH_0$ and $\tI_0$.

Specializing to 3D, we follow the same prescription as before to decompose Hamiltonians.
\begin{align}
	\tI_3 = \tI_0 \times \tI_1^3 \times \tI_2^3 \times \tI_3
\end{align}

The homotopy group $\pi_0(\tI_3) = \pi_1(\tH_2, \tI_2)$ can be computed,
\begin{align}
	\pi_1(\tI_2)
		\xrightarrow{i^2_\ast} \ColTwo{ \pi_1(\tH_2) }{0}
		\xrightarrow{j^2_\ast} \pi_0(\tI_3)
		\xrightarrow{\partial} \ColTwo{ \pi_0(\tI_2) }{\ZZ^2}
		\xrightarrow{i^2_\ast} \ColTwo{ \pi_0(\tH_2) }{\ZZ} .
\end{align}
We know that $\pi_0(\tI_2) = \ZZ^2$ from the previous section and $\pi_0(\tH_2) = \pi_2(\tH_0) = \ZZ$ corresponds to the Chern number $\tilde{G}$.
In addition, the homotopy group $\pi_2(\tH_2) = \pi_3(\tH_0) = 0$ is trivial, so $\pi_1(\tI_2)$ is irrelevant to the problem.

The map $i^2_{(0)}: \pi_0(\tI_2) \rightarrow \pi_0(\tH_2)$ is given by \eqref{eq:ChernAlphaBeta} which is surjective, hence $\coker(i^2_{(0)}) = \ZZ$.
It follows that $\pi_0(\tI_3) = \ZZ$; 3D insulators are described by the integer $\alpha_{xyz}$.
To relate $\alpha_{xyz}$ to bulk quantities, we examine the maps $\partial$ and $i^2_\ast$ under further scrutiny.
The relevant part of the exact sequence is:
\begin{align}
	0 \rightarrow \ColTwo{\pi_0(\tI_3)}{\ZZ[\alpha_{xyz}]}
		\xrightarrow{\partial} \ColTwo{ \pi_0(\tI_2) }{\ZZ^2[\alpha_{yz},\beta_{yz}]}
		\xrightarrow{i^2_{(0)}} \ColTwo{ \pi_0(\tH_2) }{\ZZ[\tilde{G}]} ,
\end{align}
where the map $i^2_{(0)}$ is given by $\tilde{G} = \alpha_{yz} + 2\beta_{yz}$.
The kernel of the map is the set $(\alpha_{yz}, \beta_{yz}) = (2j, -j)$ for integers $j$. Since the kernel is isomorphic
to $\pi_0(\tI_3)$, we can identify $j$ as $\alpha_{xyz}$; hence 
$\partial$ is defined by $\partial\alpha_{xyz}=(\alpha_{yz}, \beta_{yz}) = (2\alpha_{xyz},-\alpha_{xyz})$.

In terms of the band structure invariants, we have
\begin{align*}
	2\alpha_{xyz} & = \alpha_{yz} \big|_{k_x=\pi} \\
                      & = n_o(\pi,\pi,\pi);
\end{align*}
note that all other $n_o$'s are $0$ because the Hamiltonian is normalized.
Explicitly, the generator $H_\alpha^3$ is
\begin{align}
	H_\alpha^3 = \big[ t_0 \tau^z + \tau^x (\mathbf{t} \cdot \bm\sigma) \big]_{(++--)}
		\ominus 2 H_n^0
	,	\label{eq:GeneratorAlpha3}
\end{align}
where
\begin{align*}
		(t_0,t_x,t_y,t_z) & = \frac{(1 - x^2 - y^2 - z^2, \,2y, \,2z, \,2x)}{1 + x^2 + y^2 + z^2} ,
	\\	\textrm{with } & x = \cot\tfrac{k_x}{2}, y = \cot\tfrac{k_y}{2}, z = \cot\tfrac{k_z}{2}.
\end{align*}
At $k_x=k_y=k_z=\pi$, $(t_0,t_x,t_y,t_z)=(1,0,0,0)$ and there are two
filled bands with odd parity.
At the plane $k_x=\pi$ the Hamiltonian reduces to two copies of
\eqref{eq:GeneratorAlpha2}, but with opposite Chern numbers so that the net Chern number is $0$.  One can
think of the two subspaces of $\partial H_\alpha^3$ as $H_\alpha^2$ (Chern
$+1$) and $H_\alpha^2 \ominus H_\beta^2$ (Chern $-1$).

Therefore, there are 12 $\ZZ$ invariants in 3D.
They relate to the bulk quanties as follows.
\begin{itemize}
	\item	$n$ is the number of filled bands.
	\item	$\alpha = n_o(0,0,0)$.
	\item
		$\alpha_x = n_o(\pi,0,0)-n_o(0,0,0)$,\\
		$\alpha_y = n_o(0,\pi,0)-n_o(0,0,0)$,\\
		$\alpha_z = n_o(0,0,\pi)-n_o(0,0,0)$.
	\item
		$\alpha_{yz} = n_o(0,\pi,\pi) - \alpha_y - \alpha_z - \alpha$,\\
		$\alpha_{zx} = n_o(\pi,0,\pi) - \alpha_z - \alpha_x - \alpha$,\\
		$\alpha_{xy} = n_o(\pi,\pi,0) - \alpha_x - \alpha_y - \alpha$.
	\item
		$\beta_{yz} = \tfrac{1}{2} (\tilde{G}^{yz} - \alpha_{yz})$,\\
		$\beta_{zx} = \tfrac{1}{2} (\tilde{G}^{zx} - \alpha_{zx})$,\\
		$\beta_{xy} = \tfrac{1}{2} (\tilde{G}^{xy} - \alpha_{xy})$.
	\item
		$\alpha_{xyz} = \tfrac{1}{2} \big( n_o(\pi,\pi,\pi) - \sum \alpha_{\mu\nu} - \sum \alpha_\mu - \alpha \big)$.
\end{itemize}
The last equation explicitly written out as a function of $n_o(\bm{\kappa})$ is
\begin{align*}
	2\alpha_{xyz} & = n_o(\pi,\pi,\pi)
		\\	&\; - n_o(0,\pi,\pi) - n_o(\pi,0,\pi) - n_o(\pi,\pi,0)
		\\	&\; + n_o(\pi,0,0) + n_o(0,\pi,0) + n_o(0,0,\pi)
		\\	&\; - n_o(0,0,0) .
\end{align*}
From the formula, it is clear that the sum of parities of $n_o(\bm{\kappa})$ at the eight TRIM is even,
Eq. (\ref{eq:constraint}).

In every higher dimension $d$, $\tI_d$ has a $\ZZ$ invarant corresponding to the inversion parity $n_o(\pi,\ldots,\pi)$ generated by $H_\alpha^d$.
For the even dimensions $d = 2s$, there is a second 
$\ZZ$ invariant corresponding to the $s$\textsuperscript{th} Chern class, 
generated by $H_\beta^d$.  Hence $\pi_0(\tI_{2s}) = \ZZ^2$ 
and $\pi_0(\tI_{2s+1}) = \ZZ$.

The generator for the $s$\textsuperscript{th} Chern class is as follows.
Let $\{\Gamma^1,\Gamma^2,\ldots,\Gamma^{2s+1}\}$ be $2^s \times 2^s$ gamma matrices satisfying the Clifford algebra $\Gamma^i\Gamma^j + \Gamma^j\Gamma^i = 2\delta^{ij}$.
First we warp the Brillouin zone to a sphere: $T^{2s} [k_1,\ldots,k_{2s}] \rightarrow S^{2s} [\bm{\hat{n}}]$ in an inversion-symmetric way by sending $\vk = (\pi,\dots,\pi)$ to $\hat{n}=(1,0,0,\dots)$ and all the planes bounding
the Brillouin zone ($k_i=0$) to $(-1,0,0,\dots)$.
The Hamiltonian is defined as
\begin{align}
	H_c^{2s}(\vk) = \bm{\hat{n}} \cdot \bm\Gamma ,
\end{align}
where $\bm\Gamma$ is the $(2s+1)$-vector of gamma matrices.
\eqref{eq:GeneratorAlpha2} is an example of this construction for $d = 2s = 2$.
The inversion matrix is given by the first gamma matrix: $I_0 = \Gamma^1$, and we can see that there are $2^{s-1}$ odd-parity occupied states at $\vk=(\pi,\dots,\pi)$.  At the other TRIMs, the occupied states have even inversion-parity.

The $s$\textsuperscript{th} Chern number $C_s$ may be computed by the formula:
\begin{align}
	C_s & = \frac{1}{s!} \left(\frac{i}{2\pi}\right)^s \int\! \Tr [ P (dP)^{2s} ] ,
\end{align}
where $P(\vk) = \frac{1}{2}(1-H(\vk))$ is the projector onto the filled states and $d$ is the exterior derivative in the Brillouin zone.
Evaluating the integral shows that $C_s = \pm1$ for the 
Hamiltonian $H_c^{2s}$.

In $2s$ dimensions, Eq. (\ref{eq:2dexactseq}), generalized to more
dimensions, gives a preliminary way of choosing
the generators:
$H_\beta^{2s}$ is the image under $j$ of a generator of 
$\pi_0(\tilde{\mathcal{H}}_{2s})$
and $H_\alpha^{2s}$ is an arbitrary preimage under
$\partial^{-1}$ of the generator of $\pi_0(\tilde{\mathcal{I}})_{2s-1}$.
Any Hamiltonian can be expanded as
\begin{align}
	H = \beta_{2s} H_\beta^{2s} \oplus \alpha_{2s} H_\alpha^{2s}. \label{eq:alphabeta}
\end{align}

 $H_c^{2s}$ can be used for the generator $H_\alpha^{2s}$.
To see this, we
decompose $H_c^{2s}$ in terms of the original pair of
generators $H_\alpha^{2s}, H_\beta^{2s}$.
The $s$\textsuperscript{th} Chern number of $H_\beta^{2s}$ is $2$, since it is constructed using the map $j$ which takes a general insulator to an 
inversion-symmetric one by duplicating it in each half of the Brillouin zone [see Eq.~\eqref{eq:MapDef_j11}].
At all the TRIMs, $H_\beta^{2s}(\bm{\kappa}) = v_0$ and so $n_o$ is zero.
Using Eq. (\ref{eq:alphabeta}) for $H_c^{2s}$ implies
\begin{subequations}\begin{align}
		C_s(H_c^{2s}) = 1 &= 2\beta_{2s} + \alpha_{2s} C_s(H_\alpha^{2s}) ,
	\\	n_o(H_c^{2s}) = 2^{s-1} &= 0 + \alpha_{2s} n_o(H_\alpha^{2s}) .
\end{align}\end{subequations}
The first expression requires $\alpha_{2s}$ to be odd, and the second requires 
it to be a factor of $2^{s-1}$.
Hence $\alpha_{2s} = \pm1$ and we can use $H_c^{2s}$ as the generator $H_\alpha^{2s}$.
Since every Hamiltonian can be expressed by Eq. (\ref{eq:alphabeta}), 
the number of odd inversion-parity states $n_o(\pi,\pi,\dots)$ is 
always a multiple of $2^{s-1}$. 

In terms of the general
Hamiltonians in $\Ical_{2s}$, the total number of
odd parity states at all the TRIMs must be a multiple of $2^{s-1}$.
Furthermore,
\begin{align}
	\frac{1}{2^{s-1}} \sum_{\textrm{TRIM }\vk} n_o(\vk) \equiv C_s \pmod{2}
		&&\textrm{in $2s$-dimensions.}
\end{align}
In $2s+1$-dimensions, the Hamiltonians at $k_x = 0$ and $\pi$ are both $2s$-dimensional inversion-symmetric Hamiltonians, and they must
have the same Chern number, so the constraint on the parities is
\begin{align}
	& \sum_{\textrm{TRIM }\bm{\kappa}} n_o(\bm{\kappa}) \equiv 0 \pmod{2^s}
		&&\textrm{in $(2s+1)$-dimensions.}
		\label{eq:ChernEvenDim}
\end{align}
This is related to the ($2s+1$)-dimensional Chern-Simons integral 
$\theta_{2s+1} \in [0,2\pi)$ by
\begin{align}
	\frac{\theta_{2s+1}}{\pi} = \frac{1}{2^s} \sum_{\textrm{TRIM }\bm{\kappa}} n_o(\bm{\kappa}) \pmod{2} ,
		\label{eq:ChernOddDim}
\end{align}
because we can evaluate $\theta$ by writing the Hamiltonian as 
the $k_x = \pi$ cross-section of a Hamiltonian in one more dimension,
and then determining the Chern number of that Hamiltonian\cite{pi}.
The parity is
given by \eqref{eq:ChernEvenDim}.

\section{Tight
Binding Models and Inversion Parity.\label{app:q7}}

A useful example to study is a tight-binding
model with one atom per unit cell which has some orbitals
that are even, $|s_i\rangle$, where $i$ ranges from
$1$ to $N_{even}$ and some odd ones, $|p_i\rangle$, where
$i$ continues from $N_{even}+1$ to $N_{odd}+N_{even}$.
Let the Hamiltonian
be
\begin{equation}
H=-\sum_{i,j,\mathbf{R},\mathbf{\Delta}}t_{ij}(\bm{\Delta})c_i^\dagger(\mathbf{R}+\mathbf{\Delta}) c_j(\mathbf{R}),
\label{eq:soap}
\end{equation}
where $c_i$ includes creation operators for both types of orbitals.

A Bloch wave function $\psi_{a\mathbf{k}}$ ($a$ is the
band index)
is determined by its values at
the origin, $\phi_i=\langle s_i/p_i|\psi_{a\mathbf{k}}\rangle$, 
for short) which are repeated in a plane-wave fashion to other sites.
Inversion maps
 $\mathbf{k}\rightarrow -\mathbf{k}$ and $\phi_i\rightarrow I_{0ij}\phi_j$, 
where $I_{0ij}$ is a diagonal matrix of $N_{even}$ ones
and $N_{odd}$ minus ones. 

There is a simple criterion for determining the parities of
the occupied states at the TRIMs: if a state is odd, then all its $s$-wave
components vanish.  If it is even, all $p$-wave components vanish.
In the generic case, it is enough to check a single
$s$-wave component:
\begin{eqnarray}
\eta_{a}(\bm{\kappa})=+1 \mathrm{\ if\ }\langle s_1|\psi_{a\bm{\kappa}}\rangle\neq0\nonumber\\
\eta_{a}(\bm{\kappa})=-1 \mathrm{\ (probably)\ if\ } \langle s_1|\psi_{a\bm{\kappa}}\rangle=0.
\label{Probably}
\end{eqnarray}
If a component is exactly zero, it is reasonable to assume that it is
because
inversion symmetry
forbids that component.

The Hamiltonian used to illustrate the entanglement-rule in Fig. \ref{fig:clothchair} is of this type. In momentum space Eq. (\ref{eq:soap}) can
be written
$\sum_{ij} c_i^\dagger(\mathbf{k}) c_j(\mathbf{k}) H_{ij}(\mathbf{k})$.
The Hamiltonian 
can be built
from two subsystems $H_1,H_2$, whose occupied states are given
by the upper and lower signs in
Fig. \ref{fig:clothchair}. 
Each Hamiltonian
can be gapped because
both sets of signs 
satisfy the constraint $\prod_{\bm{\kappa}}\eta_{\bm{\kappa}}$ by itself.

Each Hamiltonian $H_r$ can be constructed in a space with one even state
$|s_r\rangle$ and one odd state $|p_r\rangle$. 
For the Hamiltonian to be inversion symmetric it must commute with $\sigma_z$,
so the diagonal matrix elements are even and the off-diagonal
matrix elements are odd functions of $\mathbf{k}$.

The two Hamiltonians can be combined together by including a small mixing
between the two systems. The final
Hamiltonian has the form:
\begin{equation}
H(\mathbf{k})=\left(\begin{array}{cc}
H_1(\mathbf{k}) & t\mathds{1}\\
t\mathds{1} & H_2(\mathbf{k})\end{array}\right).
\end{equation}
where
\begin{eqnarray}
&&H_1=A_{\mathbf{k}}\sigma_z+B_\mathbf{k}\sigma_x+C_\mathbf{k}\sigma_y\nonumber\\
&&H_2=D_{\mathbf{k}}\sigma_z+E_\mathbf{k}\sigma_x+F_\mathbf{k}\sigma_y
\end{eqnarray}
and
\begin{eqnarray*}
&&A_{\mathbf{k}}=\left(1-\cos\left(k_{x}\right)\right)\cos\left(k_{y}\right)+\left(1+\cos\left(k_{x}\right)\right)\cos\left(k_{y}-k_{z}\right)\\
&&B_{\mathbf{k}}=\left(1-\cos\left(k_{x}\right)\right)\sin\left(k_{y}\right)+\left(1+\cos\left(k_{x}\right)\right)\sin\left(k_{y}-k_{z}\right)\\
&&C_{\mathbf{k}}=\sin\left(k_{x}\right),E_{\mathbf{k}}=-\sin\left(k_{x}\right),F_{\mathbf{k}}=\sin\left(k_{z}\right)\\
&&D_{\mathbf{k}}=\cos\left(k_{x}\right)+\cos\left(k_{z}\right)-1.\end{eqnarray*}
The parities of the occupied states at $\bm{\kappa}$
of $H_1$ and $H_2$ under inversion in
the point $\frac{1}{2}\bm{\hat{x}}$ are equal to the signs
shown in Fig. \ref{fig:clothchair}.
Choosing this point to define the inversion parities
is necessary
because the entanglement spectrum is obtained
by cutting the system between two planes of the atoms.

\section{Monopoles\label{app:monopoles}}
Consider a Hamiltonian for which the $a^\mathrm{th}$ and $a+1^{\mathrm{st}}$
bands are close to being degenerate
at $\mathbf{k}_0$.  Then all states except
the two nearly degenerate states are unimportant,
and the spectrum can be described by a $2\times 2$ matrix
\begin{equation}
\mathcal{H}(\mathbf{k})
=\left(\begin{array}{cc} E_0+A(\mathbf{k})& B(\mathbf{k})\\B^*(\mathbf{k})
&E_0+C(\mathbf{k})
\end{array}\right).
\label{eq:4inone}
\end{equation}
The eigenvalues are degenerate wherever $A(\mathbf{k})=C(\mathbf{k})$ 
and $B(\mathbf{k})=0$.  This gives
three equations in three variables (since $B$ is complex), so
generically, a solution may be found in three dimensional space,
as pointed out by Von Neumann and Wigner.

Now at the location of the degeneracy (which we call $\mathbf{k}_{d}$)
there is a magnetic monopole in the Berry magnetic fields $\mathbf{B}_a$
and $\mathbf{B}_{a+1}$ of 
the two bands. The degeneracy point (or Dirac point) has
a handedness $\delta=\pm 1$, which determines the charge of the monopoles.
Bands $a$ and $a+1$ have opposite monopoles, of charge 
\begin{eqnarray}
&&Q_{a}=2\pi \delta\nonumber\\ 
&&Q_{a+1}=-2\pi \delta\label{eq:antisun}.
\end{eqnarray}
 This monopole and antimonopole are glued together
(at the point $\mathbf{k}_d$) but they do not annihilate because
they are monopoles in different magnetic fields.

To understand why the monopoles exist, write the Hamiltonian
in terms of Pauli matrices, then expand the coefficients to
linear order in $\mathbf{k}$, 
$H(\mathbf{k})\approx E_0+\mathbf{v}_0\cdot{\mathbf{k}}+\sum_{i=1}^3(\mathbf{v}_i
\cdot{\mathbf{k}}\sigma_i).$ The handedness is defined by
\begin{equation}
\delta=\mathrm{sign}[\mathbf{v}_1\cdot(\mathbf{v}_2\times\mathbf{v}_3)].
\end{equation}
Define the oblique coordinate system
$K_i=(\mathbf{k}-\mathbf{k}_d)\cdot\mathbf{v}_i$.
Now change to spherical coordinates,
$K_3=K\cos\theta, K_1=K\sin\theta\cos\phi,K_2=K\sin\theta\sin\phi$.
The eigenvalues are $E_0+\mathbf{v}_0\cdot\mathbf{k}\pm K$.
Note that if $\mathbf{v}_0=0$, then the graph of the energy
as a function of $K_1,K_2$ with $K_3$ set to $0$ is a cone.
In accordance with the degrees-of-freedom-argument given above, fixing
$K_3$ at a random value (and thus reducing the number
of parameters to 2) gives a hyperboloid, without any touching between
the bands.

The Berry magnetic field in band $a$ is defined 
as $\mathbf{B}_a=\mathrm{curl} \mathbf{A}_a$
where $\mathbf{A}_a$ is defined in Eq. (\ref{eq:strawberry});
note that the distinction between $u$ and $\psi$ does not matter for the
purpose of calculating the magnetic field.
The eigenvectors are $\psi_{a+1}=\left(\begin{array}{c}
\cos\frac{1}{2}\theta e^{-i\phi}\\\sin\frac{1}{2}\theta\end{array}\right)$
and $\psi_a=\left(\begin{array}{c}
\sin\frac{1}{2}\theta e^{-i\phi}\\-\cos\frac{1}{2}\theta\end{array}\right)$.
The Berry connections for these states are $A_{a+1}(\mathbf{K})=
\frac{1}{2}\cot\frac{\theta}{2}\hat{\phi}$ and $A_a(\mathbf{K})=
\frac{1}{2}\tan\frac{\theta}{2}\hat{\phi}$. These fields
are familiar as the vector potentials for Dirac monopoles\cite{Jackson}--
they have a flux-tube, or Dirac string, approaching
the origin along the positive and negative $K_3$ axis. These
monopoles have fluxes of $\oiint \mathbf{B}_{a+1}=-2\pi$
and $\oiint \mathbf{B}_a=2\pi$ respectively.

The calculation just completed is incorrect half the time--
the magnetic flux can be calculated in any coordinate system provided
it is right-handed; it is a pseudoscalar.
Hence, the result just obtained is correct when $\delta=+1$.
When $\delta=-1$, the oblique coordinate system is left-handed,
so the signs of the monopoles should
be flipped, leading to Eq. (\ref{eq:antisun}).

Now let us show that the curves defined by $s_a(\mathbf{k})=0$
and $s_{a+1}(\mathbf{k})=0$.  (This appears in Sec. \ref{sec:constraint}.)
The function $s_a(\mathbf{k})$ is defined as
$\langle s|\psi_{a\mathbf{k}}\rangle=\langle s|\uparrow\rangle\langle\uparrow|\psi_{a\mathbf{k}}\rangle+\langle s|\downarrow\rangle\langle\downarrow|\psi_{a\mathbf{k}}\rangle$
(since only the two states $|\uparrow\rangle$ and $|\downarrow\rangle$
that are represented by $\left(\begin{array}{c} 1\\0\end{array}\right)$
$\left(\begin{array}{c} 0\\1\end{array}\right)$ in the effective theory
are important near $\mathbf{k}_d$). So suppose
\begin{equation*}
\left(\begin{array}{c}\langle\uparrow|s\rangle \\ \langle\downarrow s\rangle
\end{array}\right)\propto
\left(\begin{array}{c} \cos\frac{\alpha}{2}e^{-i\beta}\\
\sin\frac{\alpha}{2}\end{array}\right).
\end{equation*}
 We then find
that $s^a(\mathbf{k})=0$ only if $\theta=\alpha$ and $\phi=\beta$,
and that $s^{a+1}(\mathbf{k})=0$ if $\theta=\pi-\alpha$ and $\phi=\pi+\beta$.
That is, the two curves are rays meeting at the Dirac point from
opposite directions.

Now let us see why a system cannot remain insulating when 
$\bm{n_o}$ changes.  When 
an even and odd state at a TRIM interchange, a pair of monopoles
forms or annihilates.  
This may be seen using an effective Hamiltonian near the TRIM (say
it is $\bm{\kappa}=0$).
Eq. (\ref{eq:4inone}) can be recycled for this problem. Since
$I_0H(\mathbf{k})I_0^{-1}=H(-\mathbf{k})$, and $I_0=\sigma_z$,
the diagonal entries have to be even and the off-diagonal entries
have to be odd functions of $\mathbf{k}$. 
Assume $E_0$ and
the trace are equal to zero.  (The trace just
shifts both bands by a smooth function and does not
affect the degeneracy.) We wish to study how
the dispersions change when 
$A(0)-C(0)=\Delta$, the difference between the
two energies at the TRIMs, changes sign.

Let $A(\mathbf{k})=\frac{\Delta}{2}+f(\mathbf{k})=-C(\mathbf{k})$; 
$f$ is a quadratic function to lowest order. $B(\mathbf{k})$ is a linear
function of $\mathbf{k}$, so we may choose a coordinate system where
the real and imaginary parts are $K_1$ and $K_2$,
$B(\mathbf{k})=K_1-iK_2$. 
The dispersion is therefore $\pm\sqrt{(\frac{\Delta}{2}+f(\mathbf{k}))^2+K_1^2+K_2^2}$. The states are degenerate  
if $\mathbf{K}=\mathbf{K_d}$ where $K_{d1}=K_{d2}=0$ and
$\Delta=-2f(K_{d3},0,0)=-2\alpha K_{d3}^2$, say.  
Thus, if $\alpha>0$, there
are band-touchings when $\Delta$ is negative (with
$K_{d3}=\pm\sqrt{-\frac{\Delta}{2\alpha}}$) 
and no crossings when $\Delta$ is positive.  
So if $\Delta$
changes from positive to negative, two monopoles appear. The crossings
move out from the origin to a distance proportional to
the square root of $\Delta$.

Now reexpand the Hamiltonian to linear order around $\mathbf{K}_d$.
We find that
it is equal to $\alpha K_{d3}(K_3-K_{d3})\sigma_z+K_1\sigma_x+K_2\sigma_y$.
(Notice that we
have set $K_1=K_2=0$ in $f(K_1,K_2,K_3)$ because we are interested
in values of the $K$'s such that $K_1,K_2\ll K_3$.
The orders of magnitude such that all the terms
under the square root in the dispersion have the same
magnitude are $K_1\sim \Delta$, $K_2\sim\Delta$,
$K_3\sim\sqrt{\Delta}$.)
The two cone points have opposite
handedness since $K_{d3}$ has opposite signs.  This was what we
expected
since the monopole charge has to be conserved.

\section{Entanglement Spectrum\label{app:entanglement}}
The flat band Hamiltonian is defined in terms of
the correlation function $C(x_1,x_2)=\langle \psi(x_2)^\dagger\psi(x_1)\rangle$.
(Once the results are derived in one dimension, the
three dimensional results can be written out by including factors
of 
$e^{i\mathbf{k}_\perp\cdot\mathbf{r}}$.)

 Because the correlation function decays exponentially (for an
insulator), it is reasonable to think of it as a hopping matrix for an electron system:
\begin{equation}
[H_{\mathrm{flat}}\phi](x)=\frac{1}{2}\phi(x)-\int  C(x,x')\phi(x')dx'.
\end{equation}
The eigenfunctions of this Hamiltonian are the same as the 
eigenfunctions of
the physical system but the eigenvalues are different.
Each unoccupied state 
$\phi_\gamma$ has eigenvalue $\frac{1}{2}$
and each occupied state belongs to an eigenvalue $-\frac{1}{2}$,
since $C(x_1,x_2)=\sum_\gamma \phi_\gamma(x_1)\phi_\gamma(x_2)^*$.

When a surface at $x=0$ is introduced we can
split the wave function into two parts,
$x>0$ and $x<0$, and represent them as the top and bottom halves
of state-vectors.  Then the correlation function has four parts,
\begin{equation}
\hat{C}=\left(\begin{array}{cc}\hat{C}_L & \hat{C}_{RL}^\dagger\\ \hat{C}_{RL} &\hat{C}_R \end{array}\right).
\end{equation}
The entanglement eigenstates $f^L_{i}$ turn out\cite{Peschel}
to be the eigenfunctions of $H_L=\frac{1}{2}\mathds{1}-\hat{C}_L$.

The eigenvalues are called $\frac{1}{2}-p_{i}(\mathbf{k}_\perp)$.
These eigenvalues are between $\pm\frac{1}{2}$ because $p_i$ is
the probability that an electron occupies the state $f_{i}^L$.
\emph{Any} eigenvalue in the range $0<p_{i}<1$ is a state in the gap, so the wavefunction
is localized near the surface.  There are \emph{infinitely} many surface states
like this.
In three dimensions, if they are graphed as functions of $\mathbf{k}_\perp$,
the highest bands converge to $\pm\frac{1}{2}$.

The entanglement energy spectrum is not quite the same as $p_{i}$
but is related by the transformation 
$\epsilon=2\mathrm{tanh}^{-1} (1-2p)$ which sends the limiting
energies to $\pm\infty$; $p=\frac{1}{2}$ corresponds to $\epsilon=0$.  
The graphs of the entanglement energies
look more normal--they retain a finite spacing, although there
are infinitely many of them.

Since $\hat{C}$ has $1$ and $0$ as eigenvalues,
$\hat{C}^2=\hat{C}$, giving four matrix equations.
For showing how to pair eigenstates\cite{boteroklich} 
of $\hat{C}_L$ and $\hat{C}_R$, the relevant equation 
is $\hat{C}_{RL}(\mathds{1}-\hat{C}_L)=\hat{C}_R\hat{C}_{RL}$.
Given an eigenfunction $f^L_i$ of $\hat{C}_L$ with eigenvalue $p_{Li}$
one can obtain an eigenvector of $\hat{C}_R$ with eigenvalue
$1-p_{i}$ via the transformation $\hat{M}$
\begin{equation}
f^R_i(x)=[\hat{M}f^L_i](x)=\frac{1}{\sqrt{p_i(1-p_i)}}
\sum_{x'>0} \hat{C}_{RL}(x,x')f_i^R(x').
\end{equation}
The prefactor is inserted to ensure that $f^R_i$ is normalized.
Because of how $p$ transforms, $\epsilon_{Li}=-\epsilon_{Ri}$.
(One can check that $\hat{M}$ is a unitary transformation,
which can be written in matrix form
$\hat{M}=\frac{1}{\sqrt{\hat{C}_R-\hat{C}_R^2}}\hat{C}_{RL}$).
The eigenvalues of $f^L_i$ and $f^R_i$ for $H_L$ and $H_R$ are $\pm(\frac{1}{2}-p_i)$ and $\epsilon_{Li}=-\epsilon_{Ri}$.

Furthermore, $F_i=\sqrt{p_{i}}f^L_{i}+\sqrt{1-p_{i}}f^{R}_{i}$ is
an occupied state because it satisfies $\hat{C}|F_i\rangle=|F_i\rangle$.
(Eq. (\ref{Fff}) restates this obscurely.)
States of this form give a basis for all the occupied states.
The ground state is therefore given by 
$\prod_i(\sqrt{p_i}l_i^\dagger+\sqrt{1-p_i}r_i^\dagger)|vac\rangle$
where $l_i^\dagger,r_i^\dagger$ create the states $f^L_{i}$ and $f^R_{i}$
respectively.  Cross-multiplying gives the Schmidt
decomposition, Eq. (\ref{eq:schmidt}).  The state
$F_{i}$ is definitely occupied
by an electron. This electron is on the left with probability
$p_{i}$ and on the right with probability $1-p_{i}$.  A term
in the Schmidt decomposition is obtained by making a choice,
for each mode, whether the electron is on the left or right.

If a state is filled on the left side, then the corresponding
state must be left empty
on the right.
Thus the Schmidt states are $|\alpha\rangle_L=\prod_{i\in A} l_i^{\dagger}|vac\rangle_L$
and $|\alpha\rangle_R=\prod_{i\not\in A} r_i^\dagger|vac\rangle_R$ where $A$ is a
set of states. The weight of
this state is $\prod_{i\in A} p_i\prod_{i\not\in A} (1-p_i).$

To reinterpret the fluctuations as statistical fluctuations
of the system on the left, imagine covering
the right half. Then
electrons disappear when they cross the boundary. The factors of $r_i^\dagger$
correspond to holes in $|\alpha\rangle_L$.
The weight of a state can be written in terms of just
the occupied states on the left by factoring out 
$\frac{1}{Z}:=\prod_i (1-p_i)$.  The weight is then 
$\prod_{i\in A}\frac{p_i}{1-p_i}$
which looks like a Boltzmann distribution
if $e^{-\epsilon_{Li}}:=\frac{p_i}{1-p_i}$, explaining
where the definition of $\epsilon_{Li}$ comes from.

The maximum weight Schmidt state is obtained by placing electrons
on the left half when $p_{i}>\frac{1}{2}$ and on the right
half when $p_i<\frac{1}{2}$. If we can only see the left half,
this is equivalent to filling all the negative ``energy" states.
The negative energy states on the right are also
filled since $\epsilon_{Ri}=-\epsilon_{Li}$.
So the maximum weight state is the product $|G\rangle_L|G\rangle_R$
of the ground states of $H_L$ and $H_R$.

\section{Parity of Arcs through a TRIM\label{app:fishbird}}
Suppose $\Delta N_e(\bm{\kappa}_\perp)$ is prescribed at a certain
TRIM.  The number of arcs passing through this TRIM must be equal
to $\Delta\nu$ modulo 2.
To show this, we use the $k\cdot p$ effective
Hamiltonian in the space of states that have energy zero 
at $\bm{\kappa}_\perp$
to determine
how the energies vary away from $\bm{\kappa}_\perp$. Suppose
for simplicity that all the modes at the TRIM
have the same $\mathcal{I}_e$ parity (the generic case). Then particle-hole
symmetry implies
that the effective Hamiltonian is odd
in $\mathbf{\mathbf{k}}_\perp$, to leading order $H(\mathbf{k}_\perp)=A_xk_x+A_yk_y+\dots$.
Hence $H(\mathbf{k}_\perp)=|k|(A_x\cos\theta+A_y\sin\theta)$, in polar coordinates.
The energy-dependence has a cone-like structure:
$\epsilon_i(k,\theta)=|k| f_i(\theta)$ where $f_i(\theta)$ are
the
eigenvalues of the periodic Hamiltonian $A_x\cos\theta+A_y\sin\theta$.
Now the particle-hole symmetry implies
that the bands come in pairs satisfying $\epsilon_i(\mathbf{k}_\perp)
=-\epsilon_{i'}(-\mathbf{k}_\perp)$, 
or in other words that 
$f_i(\theta+\pi)=-f_{\Delta N_e+1-i}(\theta)$.
Thus between
$\theta$ and $\theta+\pi$, the energies must be turned upside down.
This relates
the dispersions $f_i$ in pairs, except for the
middle one $f_{\frac{1}{2}(\Delta N_e+1)}$ (if $\Delta N_e$ is odd)
which is related to itself. This mode changes sign from $0$ to
$\pi$ by the symmetry, so it crosses
through 0 at an odd number $2k+1$ of values of $\theta$ 
in between, and crosses zero again at $2k+1$ points
$180^\circ$ away.  When the solutions to $\epsilon_i(\mathbf{k}_\perp)=0$
are graphed, these crossings correspond to $2k+1$ arcs
passing through the TRIM.
The other pairs of
modes $f_i$ and $f_{N+1-i}$ together give rise to an even number
of arcs. Thus, the parity of the number of Fermi arcs is
equal to the parity of $\Delta N_e$.  

Though the parity of the number of arcs is
determined by $\Delta N_e$'s parity, the precise number of
arcs is not. 
For example, when $\Delta N_e=2$, there may be no
zero-energy states away from the TRIM, as in the Dirac equation.

\section{Polarization\label{app:polarization}}

From the point of view of standard electrostatics, 
one expects a cubic crystal, aligned
with the $x,y$, and $z$ axes, with polarization $\mathbf{P}$,
to have a surface charge of $P^x$ per unit cell on the $yz$ surfaces.
However, in general, a
material may have some stray charges on the surface.  

In some situations, the surface charge is expected to be determined
(almost)
by the theoretical value of the polarization\cite{v-andk-s}.  
This does not always happen because there 
may be extra charge trapped on the surface.  But
for a \emph{clean} surface (that is, a perfectly periodic one) the 
ambiguity can be reduced: the charge density
per unit cell is given by $P^x$ 
up to an integer multiple of $e$, if the surface is gapped.

If there are no modes at the Fermi energy on the surface,
then $P^x=Q_x+ke$ (the surface charge per unit cell).
The
surface charge may
be changed by $ke$ by filling $k$ surface bands.  
It is not possible to add a fractional
multiple of $e$ to each cell, since then the surface will become
conducting,
according to the theorem that an insulator must
have a whole number of electrons per unit cell, unless
strong interactions produce an unusual phase.
(Since we are
assuming time reversal to be broken, generically, an odd number
of electrons \emph{can} form an insulator.)

This prediction for the surface
charge may be generalized to allow for a metallic surface.
In this case there is a two dimensional Fermi surface describing the modes
on the surface. The polarization is
\begin{equation}
P^x\equiv Q_x-\frac{eA_{fs}}{(2\pi)^2} \ (\mathrm{mod}\ e).
\label{juniper}
\end{equation}
where $A_{fs}$ is the area of the surface arcs.  The second term
can be interpreted as the charge that needs to be removed to make
the surface insulating.

If a crystal has a fractional polarization of $P^x$, then
there are three possibilities. The surface may
be electrically charged (with a density of $P^x+ke$ per unit cell for
some integer $k$), or it may
be metallic.  The surface may also reconstruct, so that
there is a charge density wave.  When $P^x$ is a simple
fraction, this is very likely, since
then the charge density wave would be commensurate. For example,
when $P^x=\frac{e}{2}$ (as expected for inversion-symmetric insulators)
the surface could have a period two reconstruction. The
surface might also be metallic, but a big spontaneous
surface charge seems unlikely.
Since the ``intrinsic" polarization may be exchanged
for a surface property, a scanning tunneling
microscope may be the best tool for observing it.

The polarization may be determined by noting that
the entanglement Hamiltonian $H_L$
has to satisfy the same constraints on its surface charge.
We can just determine $Q_x$ and $A_{fs}$ for the ground state
of $H_L$.
Assume $\mathbf{G}_H=0$.
Then the number of arcs through each TRIM has the same
parity, either even or odd (according to Eq. (\ref{eq:weaves})).
The
Fermi surface covers half the Brillouin zone if
this number is odd, so 
according to Eq. (\ref{eq:half}),
\begin{eqnarray}
\frac{A_{fs}}{(2\pi)^2}
&\equiv &\frac{1}{4}(\Delta N(0,0,0)+\Delta N(\pi,0,0))\nonumber\\
&\equiv &\frac{n}{2}-\tilde{P}^x_e\ (\mathrm{mod\ }1)\label{eq:nnn}
\end{eqnarray}
where $\tilde{P}^x_e$ is defined by Eq. (\ref{eq:seesaw}).
The second line uses $\Delta N(\bm{\kappa})=n-2n_o(\bm{\kappa})$

Now, if there are no nuclei at $x=0$,
then as in Sec. \ref{sec:entanglement}, one can use the neutrality
of the Schmidt spectrum and symmetry to show that $Q_x=0$.
If on $x=0$ there
are nuclei of total atomic
number per cell $Z_0$, imagine taking
these nuclei out of the system. This leaves behind
a charge of $Z_0e$ which must be divided evenly
between the left and right half. Focus
on the left half of the system; it has
a charge of $Q_x=\frac{Z_0e}{2}$.

Now substitute $Q_x$ and $A_{fs}$ into Eq. (\ref{juniper}).
Combine the $n$ term in Eq. (\ref{eq:nnn}) with $Q_x$
to get 
$\frac{(Z_0-n)e}{2}$. By neutrality $n$ is the total
atomic number per unit cell, so
$n-Z_0$ is the atomic number
of the nuclei not on $x=0$, which is congruent mod $2$
to $Z_{\frac{1}{2}}$ (the number of nuclei at $x=\frac{1}{2}$)
because the other nuclei come in pairs.
Hence the polarization is $e\tilde{P_e}^x-\frac{1}{2}Z_{\frac{1}{2}}e$.
This agrees with Eq. (\ref{eq:polarization}) because the dipole moment
of the nuclei relative to $x=0$ is $-\frac{1}{2}Z_{\frac{1}{2}}e$.
Nuclei not on the two special planes $0$ and $\frac{1}{2}$ come
in pairs and cancel out.
(We are defining the dipole moment of the nuclei
to be the sum over a
unit cell bounded by $-\frac{1}{2}<x\leq \frac{1}{2}$
in the $x$-direction.)

\section{Frozen Crystals\label{app:inertia}}
This appendix determines  what subset of the space of $\mathbf{n_o}$
vectors is spanned by integer combinations of frozen crystals.
Let $\mathbf{f}_{\mathbf{p}}$ be vectors corresponding
to systems with a single fixed electron in each unit cell
displace by $\mathbf{p}$ from the Bravais lattice.

For each of the $7$ nonzero polarizations $\mathbf{p}$, $\mathbf{f}_{\mathbf{p}}$ is
represented by a vector with four ones and four zeros ($\mathbf{f}_\mathbf{p}(\bm{\kappa})\equiv \frac{\bm{\kappa}}{\pi}\cdot\mathbf{p} \mbox{(mod 2)}$).
This assumes the electron to be in an even orbital.
For $\mathbf{p}=0$, take the electron to be in an
odd orbital instead so that
$\mathbf{f}_0(\bm{\kappa})=1$.

The goal is now to determine what vectors are integer
linear combinations of the $\mathbf{f}$'s. There
is a coordinate system for $\mathbb{Z}^8$ where this is easy to solve
One has to find a set of
vectors $\mathbf{v}_1\dots\mathbf{v}_8$ such that it
is a basis for $\mathbb{Z}^8$, and also
$n_1\mathbf{v}_1,n_2\mathbf{v}_2,\dots,n_8\mathbf{v}_8$ is
a basis for the frozen vectors (where $n_1,\dots n_8$ are certain integers).
Then if a vector $\mathbf{n_o}$ is represented by 
$a_1\mathbf{v}_1+\dots+a_8\mathbf{v}_8$ in the new coordinate system,
the criteria that it is a frozen vector are simple--$a_i$ has
to be a multiple of $n_i$. The classification theorem
for finitely generated abelian groups describes an algorithm
for finding such bases.

To find the basis, take an $8\times8$ matrix whose columns are the
$\mathbf{f}$'s and do column operations.  The only operations
that are allowed are adding or subtracting multiples of one column to another
or changing the sign of a column.  These operations do not
change the lattice spanned by the $\mathbf{f}$'s.
(They can be inverted without
dividing by integers.)

This process leads to the following basis:
$\mathbf{v}_0$,
$\mathbf{v}_x$,
$2\mathbf{v}_{xy}$, and $4\mathbf{v}_{xyz}$
and vectors symmetric with these.
Here,$\mathbf{v}_0$ is the vector with ones at all
corners of the cube,  $\mathbf{v}_x$ is the vector with ones on
the \emph{face} of the cube defined by $\kappa_x=\pi$ (and zeros
elsewhere),  $\mathbf{v}_{xy}$
is the vector with ones on the \emph{edge} defined by $\kappa_x=\kappa_y=\pi$,
and $\mathbf{v}_{xyz}$ is the vector with a one at the \emph{vertex}
$\kappa_x=\kappa_y=\pi$.

The procedure for changing the basis from $\mathbf{f}_0,\mathbf{f}_i,
\mathbf{f}_{ij},\mathbf{f}_{xyz}$ to $\mathbf{v}_0,\mathbf{v}_i,
2\mathbf{v}_{ij},4\mathbf{v}_{xyz}$ (where $i,j$ run over $x,y,z$)
consists of changing one of the four groups
of $\mathbf{f}$'s to $\mathbf{v}$'s at a time.
First $\mathbf{f}_0=\mathbf{v}_0,\mathbf{f}_i=\mathbf{v}_i$ so we can
just rename the sequence of vectors $\mathbf{v}_0,\mathbf{v}_i,\mathbf{f}_{ij},
\mathbf{f}_{xyz}$. 
Next, replace $\mathbf{f}_{\frac{1}{2}(\bm{\hat{x}}+\bm{\hat{y}})}$
by $-(\mathbf{f}_{\frac{1}{2}(\bm{\hat{x}}+\bm{\hat{y}})}
-\mathbf{v}_x-\mathbf{v}_y)$, which is equal to $2\mathbf{v}_{xy}$.
This is one of the column operations just
described (it changes a vector in the third set by
adding other vectors to it).  
Do the same for the other pairs
of $x,y,z$.  Last replace $\mathbf{f}_{\frac{1}{2}(\bm{\hat{x}}+\bm{\hat{y}}+\bm{\hat{z}})}$ by
\begin{equation*}
4\mathbf{v}_{xyz}=\mathbf{f}_{\frac{1}{2}(\bm{\hat{x}}+\bm{\hat{y}}+\bm{\hat{z}})}-\mathbf{v}_{x}-\mathbf{v}_y-\mathbf{v}_z+2\mathbf{v}_{xy}+2\mathbf{v}_{yz}
+2\mathbf{v}_{xz}.
\end{equation*}
Each of these relations makes the ones cancel out
except on just the right vertex or edge of the cube.

Some thought leads to a similar
demonstration that $\mathbf{v}_0,\mathbf{v}_i,\mathbf{v}_{ij}$
and $\mathbf{v}_{xyz}$ (without the factors of $2$ and $4$)
span \emph{all} combinations of $8$ integers.
Therefore any vector
$\mathbf{n_o}$
can be decomposed as
\begin{equation}
\mathbf{f}=a_0\mathbf{v}_0+\sum_i a_i\mathbf{v}_i+\sum_{i<j} a_{ij}\mathbf{v}_{ij}+a_{xyz}\mathbf{v}_{xyz}\label{eq:affair}
\end{equation}
where $a_{ij}$ is an even integer and $a_{xyz}$ is a multiple of four.
An even part of $a_{ij}$ and a multiple of 4 contained
in $a_{xyz}$ can be combined with the $a_0,a_i$-terms to form a
vector that represents an inert insulator
$\mathbf{f}$, leaving the remainder given in Eq. (\ref{eq:uvwxyz}).

The concise statement of this result is that the quotient
of $\mathbb{Z}^8$ by the span of the $\mathbf{f}$'s is 
$\mathbb{Z}_2^3\times\mathbb{Z}_4$.

\end{document}